\documentclass[reqno]{amsart}

\usepackage{booktabs} 
\usepackage{IEEEtrantools}
\usepackage{amssymb,latexsym,amsfonts,amsmath}
\usepackage{mathrsfs}
\usepackage{graphicx}
\usepackage{dsfont}

\usepackage{tikz}
\usetikzlibrary{calc,shapes,arrows}

\usepackage{algorithm}
\usepackage{algorithmic}

\usepackage{xspace}
\newcommand{\software}{\textsf{FAUST}$^{\mathsf 2}$\xspace}

\newtheorem{theorem}{Theorem}[section]

\newtheorem{proposition}[theorem]{Proposition}

\newtheorem{definition}[theorem]{Definition}
\newtheorem{example}[theorem]{Example}

\newtheorem{remark}[theorem]{Remark}
\newtheorem{assumption}{Assumption}
\numberwithin{equation}{section}

\newcommand{\R}{{\mathbb{R}}}

\newcommand{\Let}{:=}

\begin{document}

\begin{abstract}
This paper is concerned with a compositional approach for constructing both infinite (reduced-order models) and finite abstractions (a.k.a. finite Markov decision processes (MDPs)) of large-scale interconnected discrete-time stochastic systems. The proposed framework is based on the notion of \emph {stochastic simulation functions} enabling us to employ an abstract system as a substitution of the original one in the controller design process with guaranteed error bounds. In the first part of the paper, we derive sufficient small-gain type conditions for the compositional quantification of the probabilistic distance between the interconnection of stochastic control subsystems and that of their infinite abstractions. We then construct \emph{infinite} abstractions together with their corresponding stochastic simulation functions for a particular class of discrete-time nonlinear stochastic control systems. In the second part of the paper, we leverage small-gain type conditions for the compositional construction of \emph{finite} abstractions. We propose an approach to construct finite MDPs as finite abstractions of concrete models or their reduced-order versions satisfying an \emph{incremental input-to-state stability} property. We also show that for the particular class of nonlinear stochastic control systems, the aforementioned property can be readily checked by matrix inequalities. We demonstrate the effectiveness of the proposed results by applying our approaches to a fully interconnected network of $20$ nonlinear subsystems (totally $100$ dimensions). We construct finite MDPs from their reduced-order versions (together $20$ dimensions) with guaranteed error bounds on their output trajectories. We also apply the proposed results to a temperature regulation in a circular building and construct compositionally a finite abstraction of a network containing $1000$ rooms. We employ the constructed finite abstractions as substitutes to compositionally synthesize policies regulating the temperature in each room for a bounded time horizon.
\end{abstract}

\title[Compositional (In)Finite Abstractions for Large-Scale Interconnected Stochastic Systems]{Compositional (In)Finite Abstractions for Large-Scale Interconnected Stochastic Systems}

\author{Abolfazl Lavaei$^1$}
\author{Sadegh Soudjani$^2$}
\author{Majid Zamani$^{3,1}$}
\address{$^1$Department of Computer Science, Ludwig Maximilian University of Munich, Germany.}
\email{lavaei@lmu.de}
\address{$^2$School of Computing, Newcastle University, UK.}
\email{sadegh.soudjani@ncl.ac.uk}
\address{$^3$Department of Computer Science, University of Colorado Boulder, USA.}
\email{majid.zamani@colorado.edu}
\maketitle

\section{Introduction}
\emph{Decomposition} and in(finite) \emph{abstraction} play significant roles as two key tools in the analysis and control of large-scale interconnected systems.
Designing controllers to achieve complex specifications for large-scale systems is inherently difficult. One promising direction is to first employ abstractions of subsystems as a replacement of original (concrete) ones, then synthesize controllers for the abstract interconnected systems, and finally refine the controllers back (via an interface map) to concrete models. Since the mismatch between the output of the overall interconnected system and that of its abstraction is well-quantified, one can guarantee that the concrete system also satisfies the same specifications as the abstract one with guaranteed error bounds.

The computational complexity in synthesizing controllers for large-scale interconnected systems can be alleviated via abstractions in two consecutive stages. In the first phase, one can abstract the original system by a simpler one with a lower dimension (infinite abstractions). Then one can construct a finite abstraction as an approximate description of the (reduced-order) system in which each discrete state corresponds to a collection of continuous states of the (reduced-order) system. Since the final abstractions are finite, algorithmic machineries from computer science are applicable to synthesize controllers enforcing complex properties, e.g., expressed as temporal logic formulae, over concrete systems.

In the recent years, there have been several results on the compositional verification of stochastic models. Similarity relations over finite-state stochastic systems have been studied either via exact notions of probabilistic (bi)simulation relations~\cite{larsen1991bisimulation},~\cite{segala1995probabilistic}, or approximate versions~\cite{desharnais2008approximate},~\cite{d2012robust}. Compositional modelling and analysis for the safety verification of stochastic hybrid systems are investigated in~\cite{hahn2013compositional} in which random behaviour occurs only over the discrete components -- this limits them for being applied to systems with continuous probabilistic evolutions. Compositional controller synthesis for stochastic games using an assume-guarantee verification of a probabilistic finite automata is proposed in~\cite{basset2014compositional}. In addition, compositional probabilistic verification via an assume-guarantee framework based on a multi-objective probabilistic model checking is investigated in~\cite{kwiatkowska2013compositional} for finite systems, which supports the compositional verification for a range of quantitative properties.

There have been also several results on the construction of (in)finite abstractions for stochastic systems. Existing results include finite bisimilar abstractions for randomly switched stochastic systems~\cite{zamani2014approximately}, incrementally stable stochastic switched systems~\cite{zamani2015symbolic}, and stochastic control systems without discrete dynamics~\cite{zamani2014symbolic}. Infinite approximation techniques for jump-diffusion systems are also presented in~\cite{julius2009approximations}. In addition, compositional construction of infinite abstractions for jump-diffusion systems using small-gain type conditions is discussed in~\cite{zamani2016approximations}. Construction of finite abstractions for the formal verification and synthesis is initially proposed in~\cite{APLS08}. The improvement of construction algorithms in terms of the scalability is proposed in~\cite{SA13}. The formal abstraction-based policy synthesis is discussed in~\cite{tmka2013}, and an extension of such techniques to infinite horizon properties is discussed in~\cite{tkachev2011infinite}.  Recently, compositional construction of finite abstractions is presented in~\cite{SAM15} and~\cite{lavaei2018ADHS} using dynamic Bayesian networks and small-gain type conditions, respectively. Compositional construction of infinite abstractions (reduced-order models) is proposed in~\cite{lavaei2017compositional} and~\cite{lavaei2018CDCJ} using small-gain type conditions and dissipativity-type properties of subsystems and their abstractions, respectively, both for discrete-time stochastic control systems. 

Construction of infinite and finite abstractions using a notion of $\delta$-lifting relations is discussed in \cite{SIAM17} but without providing any compositionality result. Compositional infinite and finite abstractions in a unified framework via approximate probabilistic relations are proposed in~\cite{lavaeiNSV2019,lavaei2019NAHS1}. Compositional construction of finite MDPs for large-scale stochastic switched systems via small-gain and dissipativity approaches is respectively presented in~\cite{lavaei2019HSCC_J,lavaei2019LSS}. Compositional construction of finite abstractions for networks of not necessarily stabilizable stochastic systems via relaxed small-gain and dissipativity conditions is discussed in~\cite{lavaei2019ECC,lavaei2019CDC,lavaei2019NAHS}. An (in)finite abstraction-based technique for synthesis of stochastic control systems is recently studied in~\cite{Amy2019}.

Our main contribution here is to provide a compositional methodology for the construction of both infinite and finite abstractions. The proposed technique leverages sufficient small-gain type conditions to establish the compositionality results which rely on relations between subsystems and their abstractions described by the existence of so-called stochastic simulation functions. This type of relations enables us to compute the probabilistic error between the interconnection of concrete subsystems and that of their (in)finite abstractions. In this respect, we first construct infinite abstractions together with their corresponding stochastic simulation functions for a particular class of discrete-time nonlinear stochastic control systems. We then propose an approach to construct finite Markov decision processes of concrete discrete-time stochastic control systems (or their reduced-order versions) satisfying an incremental input-to-state stability property. We show that for the particular class of discrete-time nonlinear stochastic control systems, the aforementioned property can be readily checked by matrix inequalities.  

To show the applicability of our approach to strongly connected networks with nonlinear dynamics, we apply the results to a fully interconnected network of $20$ nonlinear subsystems (totally $100$ dimensions) and construct finite MDPs from their reduced-order versions (together $20$ dimensions) with guaranteed probabilistic error bounds between their output trajectories. We also apply our proposed technique to a temperature regulation in a circular building and construct compositionally a finite abstraction of a network containing $1000$ rooms. We employ the constructed finite abstractions as substitutes to compositionally synthesize policies regulating the temperature in each room for a bounded time horizon.

{\bf Related literature.} Our proposed approach here differs from the one in~\cite{lavaei2018ADHS} in three main directions. First and foremost, we provide a compositional approach here for the construction of both infinite (reduced-order models) and finite abstractions (finite MDPs) (cf. Sections~\ref{sec:constrcution_infinite},\ref{sec:compositionality1}), while the proposed compositional scheme in~\cite{lavaei2018ADHS} is only for the construction of finite abstractions. Second, we provide an approach for the construction of finite MDPs for both the general setting of nonlinear stochastic systems, and a particular class of discrete-time nonlinear stochastic control systems (cf. Subsections~\ref{subsec:general setting110} and \ref{subsec:nonlinear110}), while the construction scheme in~\cite{lavaei2018ADHS} only handles the class of linear systems. As our third contribution, we apply our results to a fully interconnected nonlinear network by constructing finite MDPs from their reduced-order versions with guaranteed error bounds. In addition, we provide the proofs of all statements which were omitted in \cite{lavaei2018ADHS}.

Compositional construction of infinite abstractions for interconnected discrete-time stochastic control systems is also proposed in \cite{lavaei2017compositional}. Although the provided results in~\cite{lavaei2017compositional} are also about infinite abstractions, the compositional framework there is based on a restrictive small-gain condition. More precisely, our compositionality results here are based on a $\max$ small-gain condition which is more general than the classic one provided in~\cite{lavaei2017compositional} since the proposed $\max$ small-gain condition does not require any linear growth on the gains of subsystems which is the case in~\cite{lavaei2017compositional} (cf. comparable Example~\ref{comparable example}). In addition, the provided approximation error in~\cite[inequality (7)]{lavaei2017compositional} increases as the number of subsystems grows. Whereas, our error provided in~\eqref{Eq_25} does not change since the overall error is completely independent of the size of the network, and is computed only based on the maximum of errors of subsystems instead of being a linear combination of them which is the case in~\cite{lavaei2017compositional}. We also provide an approach for the construction of infinite MDPs for a particular class of nonlinear stochastic systems (cf. Subsection~\ref{subsec:nonlinear}), whereas the construction scheme for infinite abstractions proposed in~\cite{lavaei2017compositional} only handles the class of linear systems.

Recently, compositional construction of finite abstractions for networks of discrete-time stochastic control systems is also proposed in \cite{lavaei2017HSCC}, but using a different compositionality scheme based on the dissipativity theory. In general, the proposed compositional synthesis approach here is much \emph{less conservative} than the one provided in \cite{lavaei2017HSCC} since the overall approximation error here is computed based on the maximum error of subsystems instead of their linear combinations which is the case in \cite{lavaei2017HSCC}. We refer the interested readers to \cite{lavaei2018ADHS} for a detailed comparison between these two compositionality schemes on a case study. 

Compositional construction of finite abstractions for discrete-time Markov processes is also proposed in~\cite{SAM15} using finite dynamic Bayesian networks. The proposed approach in~\cite{SAM15} is more general than our setting here since the proposed framework in~\cite{SAM15} does not require original systems to be incremental input-to-state stable. On the other hand, the abstraction error in~\cite{SAM15} depends on the Lipschitz constants of stochastic kernels associated with the system. This error converges to infinity when the standard deviation of the noise goes to zero which is not the case in our setting. Thus, our proposed approach outperforms significantly the results in~\cite{SAM15} for noises with a small standard deviation.

\section{Discrete-Time Stochastic Control Systems}\label{Preliminaries}

\subsection{Preliminaries}
A probability space in this work is presented by $(\Omega,\mathcal F_{\Omega},\mathbb{P}_{\Omega})$,
where $\Omega$ is a sample space,
$\mathcal F_{\Omega}$ is a sigma-algebra on $\Omega$ which comprises subsets of $\Omega$ as events,
and $\mathbb{P}_{\Omega}$ is a probability measure that assigns probabilities to events.
Random variables introduced here are measurable functions of the form $X:(\Omega,\mathcal F_{\Omega})\rightarrow (S_X,\mathcal F_X)$ such that any random variable $X$ induces a probability measure on  its space $(S_X,\mathcal F_X)$ as $Prob\{A\} = \mathbb{P}_{\Omega}\{X^{-1}(A)\}$ for any $A\in \mathcal F_X$. We directly present the probability measure on $(S_X,\mathcal F_X)$ without explicitly mentioning the underlying probability space and the function $X$ itself.

We call the topological space $S$ as a Borel space if it is homeomorphic to a Borel subset of a Polish space (i.e., a separable and completely metrizable space). Euclidean space $\mathbb R^n$, its Borel subsets endowed with a subspace topology, and hybrid spaces are examples of a Borel space. A Borel sigma-algebra is denoted by $\mathcal B(S)$, and any Borel space $S$ is assumed to be endowed with it. A map $f : S\rightarrow Y$ is measurable whenever it is Borel measurable.

\subsection{Notation}

The sets of nonnegative and positive integers are denoted by $\mathbb N := \{0,1,2,\ldots\}$ and $\mathbb N_{\ge 1} := \{1,2,3,\ldots\}$, respectively. Moreover,
the symbols $\mathbb R$, $ \mathbb R_{>0}$, and $\mathbb R_{\ge 0}$ denote, respectively, the sets of real, positive and nonnegative real numbers.
Given $N$ vectors $x_i \in \mathbb R^{n_i}$, $n_i\in \mathbb N_{\ge 1}$, and $i\in\{1,\ldots,N\}$, we use $x = [x_1;\ldots;x_N]$ to denote the corresponding vector of the dimension $\sum_i n_i$.
We denote by $\Vert\cdot\Vert$ and $\Vert\cdot\Vert_2$ the infinity and Euclidean norms, respectively. Given any $a\in\mathbb R$, $\vert a\vert$ denotes the absolute value of $a$. Symbols $\mathds{I}_n$, $\mathbf{0}_n$, and $\mathds{1}_n$ denote the identity matrix in $\mathbb R^{n\times{n}}$ and the column vector in $\mathbb R^{n\times{1}}$ with all elements equal to zero and one, respectively. The identity
function and composition of functions are denoted by $\mathcal{I}_d$ and symbol $\circ$, respectively. We denote by $\mathsf{diag}(a_1,\ldots,a_N)$ a diagonal matrix in $\mathbb R^{N\times{N}}$ with diagonal matrix entries $a_1,\ldots,a_N$ starting from the upper left corner. Given functions $f_i:X_i\rightarrow Y_i$,
for any $i\in\{1,\ldots,N\}$, their Cartesian product $\prod_{i=1}^{N}f_i:\prod_{i=1}^{N}X_i\rightarrow\prod_{i=1}^{N}Y_i$ is defined as $(\prod_{i=1}^{N}f_i)(x_1,\ldots,x_N)=[f_1(x_1);\ldots;f_N(x_N)]$.
For any set $\mathcal A$, we denote by $\mathcal A^{\mathbb N}$ the Cartesian product of a countable number of copies of $\mathcal A$, i.e., $\mathcal A^{\mathbb N} = \prod_{k=0}^{\infty} \mathcal A$. A function $\gamma:\mathbb\mathbb \mathbb R_{\ge 0}\rightarrow\mathbb\mathbb \mathbb R_{\ge 0}$, is said to be a class $\mathcal{K}$ function if it is continuous, strictly increasing, and $\gamma(0)=0$. A class $\mathcal{K}$ function $\gamma(\cdot)$ is said to be a class $\mathcal{K}_{\infty}$ if
$\gamma(r) \rightarrow \infty$ as $r\rightarrow\infty$.

\subsection{Discrete-Time Stochastic Control Systems}
In this paper, stochastic control systems in discrete time (dt-SCS) are defined
by the tuple
\begin{equation}
\label{eq:dt-SCS}
\Sigma=\left(X,U,W,\varsigma,f,Y, h\right)\!,
\end{equation}
where $X\subseteq \mathbb R^n$ is a Borel space as the state space of the system.
The measurable space with $\mathcal B (X)$  being  the Borel sigma-algebra on the state space is denoted by $(X, \mathcal B (X))$. Sets
$U\subseteq \mathbb R^m$ and $W\subseteq \mathbb R^p$ are Borel spaces as \emph{external} and \emph{internal} input spaces of the system.
Notation $\varsigma$ denotes a sequence of independent and identically distributed (i.i.d.) random variables from a sample space $\Omega$ to the set $V_\varsigma$,
\begin{equation*}
\varsigma:=\{\varsigma(k):\Omega\rightarrow V_{\varsigma},\,\,k\in\mathbb N\}.
\end{equation*}
The map $f:X\times U\times W\times V_{\varsigma} \rightarrow X$ is a measurable function characterizing the state evolution of the system. Finally, the set $Y\subseteq \mathbb R^q$ is a Borel space as the output space of the system, and the map $h:X\rightarrow Y$ is a measurable function that maps a state $x\in X$ to its output $y = h(x)$.

An evolution of the state of dt-SCS $\Sigma$ for a given initial state $x(0)\in X$ and input sequences $\nu(\cdot):\mathbb N\rightarrow U$ and $w(\cdot):\mathbb N\rightarrow W$ is described as
\begin{equation}\label{Eq_1a}
\Sigma:\left\{\hspace{-1.5mm}\begin{array}{l}x(k+1)=f(x(k),\nu(k),w(k),\varsigma(k)),\\
y(k)=h(x(k)),\\
\end{array}\right.
\quad k\in\mathbb N.
\end{equation}

A dt-SCS $\Sigma$ in \eqref{eq:dt-SCS} can be \emph{equivalently} represented as a Markov decision process (MDP) \cite{SIAM17}
\begin{equation}\notag
\Sigma=\left(X,U,W,T_{\mathsf x},Y,h\right)\!,	
\end{equation}
having a general state space $X$, where the map $T_{\mathsf x}:\mathcal B(X)\times X\times U\times W\rightarrow[0,1]$,
is a conditional stochastic kernel that assigns to any $x \in X$, $\nu\in U$, and $w\in W$, a probability measure $T_{\mathsf x}(\cdot | x,\nu, w)$
on the measurable space
$(X,\mathcal B(X))$
so that for any set $\mathcal A \in \mathcal B(X)$, 
$$\mathbb P (x(k+1)\in \mathcal A\,|\, x(k),\nu(k),w(k)) = \int_{\mathcal A} T_{\mathsf x} (d x'|x(k),\nu(k),w(k)).$$
For given inputs $\nu(\cdot), w(\cdot),$  the stochastic kernel $T_{\mathsf x}$ captures the evolution of the state of $\Sigma$ and can be uniquely determined by the pair $(\varsigma,f)$ from \eqref{eq:dt-SCS}.

We are interested in Markov policies, defined next, to control the system given the dt-SCS in~\eqref{eq:dt-SCS}. 
\begin{definition}
	For the dt-SCS $\Sigma$ in~\eqref{eq:dt-SCS}, a  Markov policy is a sequence
	$\rho = (\rho_0,\rho_1,\rho_2,\ldots)$ of universally measurable stochastic kernels $\rho_n$~\cite{Bertsekas1996}, each defined on the input space $U$ given $X\times W$ such that for all $(x_n,w_n)\in X\times W$, $\rho_n(U|(x_n,w_n))=1$.
	The class of all such Markov policies is denoted by $\Pi_M$. 
\end{definition} 

We associate to $U$ and $W$ the sets $\mathcal U$ and $\mathcal W$ respectively to be collections of sequences $\{\nu(k):\Omega\rightarrow U,\,\,k\in\mathbb N\}$ and $\{w(k):\Omega\rightarrow W,\,\,k\in\mathbb N\}$, in which $\nu(k)$ and $w(k)$ are independent of $\varsigma(t)$ for any $k,t\in\mathbb N$ and $t\ge k$.
The random sequences $x_{a\nu w}:\Omega \times\mathbb N \rightarrow X$, $y_{a\nu w}:\Omega \times \mathbb N \rightarrow Y$ satisfying~\eqref{Eq_1a} for any initial state $a\in X$, $\nu(\cdot)\in\mathcal{U}$, and $w(\cdot)\in\mathcal{W}$ are called the \textit{solution process} and the \textit{output trajectory} of $\Sigma$ respectively under an external input $\nu$, an internal input $w$, and an initial state $a$.

In this paper, our main contribution is to study the interconnected discrete-time stochastic control systems without internal signals resulting from the interconnection of dt-SCS having both internal and external signals. Then the interconnected dt-SCS without internal signal is reduced to the tuple $(X,U,\varsigma,f,Y,h)$, where $f:X\times U\times V_\varsigma\rightarrow X$.

In the next sections, we provide an approach for the compositional synthesis of (in)finite abstractions for interconnected dt-SCS. To do so, we first define the notions of stochastic pseudo-simulation and simulation functions for quantifying the error between two dt-SCS (with both internal and external signals) and two interconnected dt-SCS (without internal signals), respectively.

\section{Stochastic (Pseudo-)Simulation Functions}
\label{sec:SPSF}
In this section, for dt-SCS with both internal and external signals, we first introduce the notion of stochastic pseudo-simulation functions (SPSF).  We then define the notion of stochastic simulation functions (SSF) for dt-SCS without internal signals. Although the former definition is employed to quantify the closeness of two dt-SCS, the latter is specifically employed for the interconnected dt-SCS.
\begin{definition}\label{Def_1a}
	Consider two dt-SCS $\Sigma =(X,U,W,\varsigma,f,Y,h)$ and
	$\widehat\Sigma =(\hat X,\hat U, \hat W, \varsigma,\hat f, \hat Y, \hat h)$, where $\hat W\subseteq W$ and $\hat Y\subseteq Y$. A function $S:X\times\hat X\to\mathbb R_{\ge0}$ is
	called a stochastic pseudo-simulation function (SPSF) from  $\widehat\Sigma$ to $\Sigma$ if there exist functions
	$\alpha,\kappa\in\mathcal{K}_\infty$, with $\kappa<\mathcal{I}_d$, $\rho_{\mathrm{int}},\rho_{\mathrm{ext}}\in\mathcal{K}_\infty\cup\{0\}$, and a constant $\psi \in\mathbb R_{\ge 0}$, such that
	\begin{align}\label{Eq_2a}
	\alpha(\Vert h(x)-\hat h(\hat x)\Vert)\le S(x,\hat x),\quad\forall x\in X, \hat x\in\hat X,
	\end{align}
	and for all $x\in X,\,\hat x\in\hat X,\,\hat\nu\in\hat U$ there exists $\nu\in U$ such that $\forall \hat w\in\hat W$, $\forall w\in W$,
	\begin{align}
	\notag
	\mathbb{E}& \Big[S(f(x,\nu,w,\varsigma),\hat{f}(\hat x,\hat \nu,\hat w,\varsigma))\,\big|\,x,\hat{x}, \nu,\hat{\nu}, w,\hat w\Big]\\\label{Eq_3a}
	&\leq \max\Big\{\kappa(S(x,\hat{x})), \rho_{\mathrm{int}}(\Vert w-\hat w\Vert), 
	\rho_{\mathrm{ext}}(\Vert\hat\nu\Vert),\psi\Big\}.
	\end{align}
\end{definition}
We denote $\widehat\Sigma\preceq_{\mathcal{PS}}\Sigma$
if there exists an SPSF $S$ from $\widehat\Sigma$ to $\Sigma$, and call the control system $\widehat\Sigma$ an abstraction of the concrete (original) system $\Sigma$. Note that $\widehat{\Sigma}$ may be finite or infinite depending on cardinalities of sets $\hat X,\hat U,\hat W$.

\begin{remark}
	As a comparison, the notion of SPSF here is equivalent to the one defined in~\cite[Definition 3.1]{lavaei2017compositional} such that the existence of one implies that of the other one. However, the upper bound in \eqref{Eq_3a} is in the $\max$ form, whereas the one in~\cite[inequality (4)]{lavaei2017compositional} is in the additive form.
\end{remark}

\begin{remark}
	Second condition in Definition~\ref{Def_1a} implies implicitly the existence of an interface function $\nu=\nu_{\hat \nu}(x,\hat x,\hat \nu)$ satisfying the inequality~\eqref{Eq_3a} which can be employed to refine a synthesized policy $\hat\nu$ for $\widehat\Sigma$ to a policy $\nu$ for $\Sigma$. 
\end{remark}

Definition~\ref{Def_1a} can also be stated for systems without internal signals by eliminating all the terms related to $w,\hat w$. The precise definition is provided in Definition~\ref{Def_Com1} in Appendix.

The next theorem shows how an SSF can be employed to compare output trajectories of two interconnected dt-SCS (without internal signals) in a probabilistic sense. This theorem is borrowed from~\cite[Theorem 3.3]{lavaei2017compositional}, and holds for our setting here since the $\max$ form of SSF here implies the additive form used in \cite{lavaei2017compositional}.

\begin{theorem}\label{Thm_1a}
	Let
	$\Sigma =(X,U,\varsigma,f, Y,h)$ and
	$\widehat\Sigma =(\hat X,\hat U,\varsigma,\hat f, \hat Y,\hat h)$
	be two dt-SCS without internal signals, where $\hat Y\subseteq Y$.
	Suppose $V$ is an SSF from $\widehat\Sigma$ to $\Sigma$, and there exists a constant $0<\hat\kappa<1$ such that the function $\kappa \in \mathcal{K}_\infty$ in~\eqref{eq6666} satisfies $\kappa(s)\geq\hat\kappa s$, $\forall s\in\mathbb R_{\geq0}$. For any external input trajectory $\hat\nu\in\mathcal{\hat U}$ that preserves Markov property for the closed-loop $\widehat\Sigma$, and for any random variables $a$ and $\hat a$ as initial states of the two dt-SCS,
	there exists an input trajectory $\nu\in\mathcal{U}$ of $\Sigma$ through the interface function associated with $V$ such that the following inequality holds	
	\begin{align}\label{Eq_25}
	&\mathbb{P}\left\{\sup_{0\leq k\leq T_d}\Vert y_{a\nu}(k)-\hat y_{\hat a \hat\nu}(k)\Vert\geq\varepsilon\,|\,[a;\hat a]\right\}\leq \hat\delta,\\\notag
	&\hat\delta :=
	\begin{cases}
	1-(1-\frac{V(a,\hat a)}{\alpha\left(\varepsilon\right)})(1-\frac{\widehat\psi}{\alpha\left(\varepsilon\right)})^{T_d},&\quad\quad \text{if}~\alpha\left(\varepsilon\right)\geq\frac{\widehat\psi}{\hat\kappa},\\
	(\frac{V(a,\hat a)}{\alpha\left(\varepsilon\right)})(1-\hat\kappa)^{T_d}+(\frac{\widehat\psi}{\hat\kappa\alpha\left(\varepsilon\right)})(1-(1-\hat\kappa)^{T_d}), &\quad\quad \text{if}~\alpha\left(\varepsilon\right)<\frac{\widehat\psi}{\hat\kappa},
	\end{cases}
	\end{align}
	for any $\varepsilon>0$, where the constant $\widehat\psi\geq0$ satisfies  $\widehat\psi\geq \rho_{\mathrm{ext}}(\Vert\hat \nu\Vert_{\infty})+\psi$.
\end{theorem}

\begin{remark}
	Note that $\psi=0$ possibly if concrete and abstract systems are both continuous-space but perhaps with different dimensions and share the same multiplicative noise (cf. Eq. (2) in \cite{lavaei2017compositional}). In this case, the function $V$ becomes a nonnegative supermartingale if $\rho_{\mathrm{ext}}(\cdot)$ is also equal to zero. Then one can readily extend the result of Theorem~\ref{Thm_1a} to the infinite-time horizon and compute the mismatch between two interconnected systems by applying the results in ~\cite[Corollary 3.4]{lavaei2017compositional}.
\end{remark}
The next proposition establishes a so-called transitivity property for the computation of error bounds proposed in Theorem \ref{Thm_1a}. This result is important especially when one first constructs a reduced-order model (an infinite abstraction) of an original stochastic system and then uses it to construct a finite MDP. The next proposition can provide the overall error bound in this two-step abstraction scheme. We refer the interested readers to the first case study in Section \ref{example} for an application of this proposition.
\begin{proposition}\label{proposition}
	Suppose $\Sigma_1$, $\Sigma_2$, and $\Sigma_3$ are three dt-SCS without internal signals. For any external input trajectories $\nu_1$, $\nu_2$, and $\nu_3$ and for any random variables $a_1$, $a_2$, and $a_3$ as initial states of the three dt-SCS, if
	\begin{align}\notag
	&\mathbb{P}\left\{\sup_{0\leq k\leq T_d}\Vert y_{1a_1\nu_1}(k)-y_{2 a_2\nu_2}(k)\Vert\geq\varepsilon_1\,|\,[a_1;a_2]\right\}\leq \hat\delta_1,\\\notag
	&
	\mathbb{P}\left\{\sup_{0\leq k\leq T_d}\Vert y_{2 a_2\nu_2}(k)-y_{3a_3 \nu_3}(k)\Vert\geq\varepsilon_2\,|\,[a_2;a_3]\right\}\leq \hat\delta_2,
	\end{align}
	for some $\varepsilon_1,\varepsilon_2>0$ and $\hat\delta_1,\hat\delta_2\in]0~1[$, then the probabilistic mismatch between output trajectories of $\Sigma_1$ and $\Sigma_3$ is quantified as
	\begin{align}\notag
	\mathbb{P}&\left\{\sup_{0\leq k\leq T_d}\Vert y_{1a_1\nu_1}(k)-y_{3a_3\nu_3}(k)\Vert\geq\varepsilon_1+\varepsilon_2\,|\,[a_1;a_2;a_3]\right\}\leq \hat\delta_1+\hat\delta_2.
	\end{align}	
\end{proposition}
The proof is provided in Appendix.

\section{Interconnected Stochastic Control Systems}\label{sec:compositionality}
We consider a collection of stochastic control subsystems 
\begin{equation}
\label{eq:network}
\Sigma_i=(X_i,U_i,W_i,\varsigma_i,f_i,Y_{i},h_{i}),\quad i\in \{1,\dots,N\},
\end{equation}
where their internal inputs and outputs are partitioned as
\begin{align}\notag
w_i&=[{w_{i1};\ldots;w_{i(i-1)};w_{i(i+1)};\ldots;w_{iN}}],\\\label{config1}
y_i&=[{y_{i1};\ldots;y_{iN}}],
\end{align}
and their output spaces and functions are of the form
\begin{equation}
\label{config2}
Y_i=\prod_{j=1}^{N}Y_{ij},\quad h_i(x_i)=[{h_{i1}(x_i);\ldots;h_{iN}(x_i)}].
\end{equation}
Outputs $y_{ii}$ are interpreted as \emph{external} ones, whereas outputs $y_{ij}$ with $i\neq j$ are \emph{internal} ones which are employed to interconnect these stochastic control subsystems. For the interconnection, if there is a connection from $\Sigma_{j}$ to $\Sigma_i$, we assume that $w_{ij}$ is equal to $y_{ji}$. Otherwise, we put the connecting output function identically zero, i.e., $h_{ji}\equiv 0$. Now we define concrete interconnected stochastic control systems.

\begin{definition}\label{Concrete interconnection}
	Consider $N\in\mathbb N_{\geq1}$ stochastic control subsystems $\Sigma_i=(X_i,U_i,W_i,\varsigma_i,f_i,Y_{i},h_{i})$, $i\in \{1,\dots,N\}$, with the input-output configuration as in \eqref{config1} and \eqref{config2}. The 
	interconnection of  $\Sigma_i$ for any $i\in \{1,\ldots,N\}$, is the \emph{concrete} interconnected stochastic control system $\Sigma=(X,U,\varsigma,f,Y,h)$, denoted by
	$\mathcal{I}(\Sigma_1,\ldots,\Sigma_N)$, such that $X:=\prod_{i=1}^{N}X_i$,  $U:=\prod_{i=1}^{N}U_i$, $f:=\prod_{i=1}^{N}f_{i}$, $Y:=\prod_{i=1}^{N}Y_{ii}$, and $h=\prod_{i=1}^{N}h_{ii}$, subjected to the following constraint:
	\begin{equation*}
	\forall i,j\in \{1,\dots,N\},i\neq j\!: ~~~ w_{ji} = y_{ij}, ~~~ Y_{ij}= W_{ji}.
	\end{equation*}
\end{definition}

\begin{figure}[ht]
	\begin{tikzpicture}[>=latex']
	\tikzstyle{block} = [draw, 
	thick,
	rectangle, 
	minimum height=.8cm, 
	minimum width=1.5cm]
	
	\node at (-3.5,-0.75) {$\mathcal{I}(\Sigma_1,\Sigma_2)$};
	
	\draw[dashed] (-1.7,-2.2) rectangle (1.7,.7);
	
	\node[block] (S1) at (0,0) {$\Sigma_1$};
	\node[block] (S2) at (0,-1.5) {$\Sigma_2$};
	
	\draw[->] ($(S1.east)+(0,0.25)$) -- node[very near end,above] {$y_{11}$} ($(S1.east)+(1.5,.25)$);
	\draw[<-] ($(S1.west)+(0,0.25)$) -- node[very near end,above] {$\nu_{1}$} ($(S1.west)+(-1.5,.25)$);
	
	\draw[->] ($(S2.east)+(0,-.25)$) -- node[very near end,below] {$y_{22}$} ($(S2.east)+(1.5,-.25)$);
	\draw[<-] ($(S2.west)+(0,-.25)$) -- node[very near end,below] {$\nu_{2}$} ($(S2.west)+(-1.5,-.25)$);
	
	\draw[->] 
	($(S1.east)+(0,-.25)$) -- node[very near end,above] {$y_{12}$} 
	($(S1.east)+(.5,-.25)$) --
	($(S1.east)+(.5,-.5)$) --
	($(S2.west)+(-.5,.5)$) --
	($(S2.west)+(-.5,.25)$) -- node[very near start,below] {$w_{21}$}
	($(S2.west)+(0,.25)$) ;
	
	\draw[->] 
	($(S2.east)+(0,.25)$) -- node[very near end,below] {$y_{21}$} 
	($(S2.east)+(.5,.25)$) --
	($(S2.east)+(.5,.5)$) --
	($(S1.west)+(-.5,-.5)$) --
	($(S1.west)+(-.5,-.25)$) -- node[very near start,above] {$w_{12}$}
	($(S1.west)+(0,-.25)$) ;
	
	\end{tikzpicture}
	\caption{Interconnection of two \emph{concrete} stochastic subsystems $\Sigma_1$ and $\Sigma_2$.}
	\label{system1}
\end{figure}
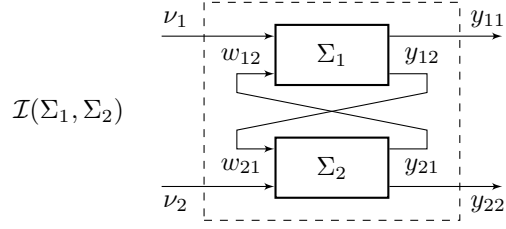

An example of the interconnection of two concrete control subsystems $\Sigma_1$ and $\Sigma_2$ is illustrated in Figure \ref{system1}.

\section{Compositional Infinite Abstractions for Interconnected Systems}\label{sec:constrcution_infinite}
In this section, we analyze networks of stochastic control subsystems and discuss how to construct their infinite abstractions together with a simulation function based on corresponding SPSF functions of their subsystems. We consider here $\Sigma$ as an original dt-SCS and $\widehat{\Sigma}$ as its infinite abstraction with (potentially) a lower dimension. Suppose we are given $N$ concrete stochastic control subsystems in \eqref{eq:network} together with their corresponding infinite abstractions
\begin{align}\label{infinite_abst}
\widehat\Sigma_i=(\hat X_i,\hat U_i,\hat W_i,\varsigma_i,\hat f_i,\hat Y_i,\hat h_i),	
\end{align}
where $\hat W_i= W_{i}$ and $\hat Y_i= Y_{i}$, with an SPSF $S_i$ from $\widehat\Sigma_i$ to $\Sigma_i$ with the corresponding functions and constants denoted by
$\alpha_i$, $\kappa_i$, $\rho_{\text{int}i}$, $\rho_{\text{ext}i}$, and $\psi_i$. Now we raise the following small-gain assumption that is essential for the compositionality result in this section.

\begin{assumption}\label{Assump: Gamma}
	Assume that $\mathcal{K}_\infty$ functions $\kappa_{ij}$ defined as
	\begin{equation*}
	\kappa_{ij}(s) := 
	\begin{cases}
	\kappa_i(s), \quad\quad& \text{if }i = j,\\
	\rho_{\mathrm{int}i}(\alpha_j^{-1}(s)), \quad\quad& \text{if }i \neq j,
	\end{cases}
	\end{equation*}
	satisfy
	\begin{equation}
	\label{Assump: Kappa}
	\kappa_{i_1i_2}\circ\kappa_{i_2i_3}\circ\dots \circ \kappa_{i_{r-1}i_{r}}\circ\kappa_{i_{r}i_1} < \mathcal{I}_d
	\end{equation}	
	for all sequences $(i_1,\dots,i_{r}) \in \{1,\dots,N\}^ {r}$ and $r\in \{1,\dots,N\}$. 
	\begin{remark}
		Note that the small-gain condition~\eqref{Assump: Kappa} is a standard one in studying the stability of large-scale interconnected systems via ISS Lyapunov functions~\cite{dashkovskiy2007iss,dashkovskiy2010small}. This condition is automatically satisfied if each $\kappa_{ii}$ is less than identity ($\kappa_{ii}<\mathcal{I}_d, \forall i\in\{1,\dots,N\}$). Although  this condition should be satisfied for all possible sequences $(i_1,\dots,i_{r}) \in \{1,\dots,N\}^ {r}, r\in \{1,\dots,N\}$, it allows some subsystems to compensate the undesirable effects of other subsystems in the interconnected network such that this condition is satisfied.
	\end{remark}
	\begin{remark}
		We emphasize that the proposed $\max$ small-gain condition~\eqref{Assump: Kappa} is more general than the \emph{classic} one provided in~\cite{lavaei2017compositional} since it does not require any linear growth on the gains of subsystems which is the case in~\cite{lavaei2017compositional}. See Example~\ref{comparable example} in Appendix for a comparison.
	\end{remark}			
	The small-gain condition~\eqref{Assump: Kappa} implies the existence of $\mathcal{K}_\infty$ functions $\sigma_i>0$ \cite[Theorem 5.5]{ruffer2010monotone}, satisfying
	\begin{align}\label{compositionality}
	\max_{i,j}\Big\{\sigma_i^{-1}\circ\kappa_{ij}\circ\sigma_j\Big\} < \mathcal{I}_d, \quad i,j = \{1,\dots,N\}.
	\end{align}
\end{assumption}	
In the next theorem, we show that if Assumption \ref{Assump: Gamma} holds and $\max_{i}\sigma_i^{-1}$ is concave (in order to employ Jensen's inequality), then we can compute the mismatch between the interconnection of stochastic control subsystems and that of their infinite abstractions in a compositional fashion.

\begin{theorem}\label{Thm: Comp}
	Consider the interconnected dt-SCS
	$\Sigma=\mathcal{I}(\Sigma_1,\ldots,\Sigma_N)$ induced by $N\in\mathbb N_{\geq1}$ stochastic
	control subsystems~$\Sigma_i$. Suppose that each $\Sigma_i$ admits an infinite abstraction $\widehat \Sigma_i$ together with a corresponding SPSF $S_i$. If Assumption~\ref{Assump: Gamma} holds and $\max_{i}\sigma_i^{-1}$ for $\sigma_i$ as in \eqref{compositionality} is concave,
	then the function $V(x,\hat x)$ defined as
	\begin{equation}
	\label{Comp: Simulation Function}
	V(x,\hat x) := \max_{i} \Big\{\sigma_i^{-1}(S_i(x_i,\hat x_i))\Big\},
	\end{equation}
	is an SSF from $\widehat \Sigma=\mathcal{I}(\widehat {\Sigma}_1,\ldots,\widehat{\Sigma}_N)$ to $\Sigma=\mathcal{I}(\Sigma_1,\ldots,\Sigma_N)$. 	
\end{theorem}

The proof of Theorem~\ref{Thm: Comp} is provided in Appendix. 

Now in the next subsection, we propose an approach to construct infinite abstractions for a particular class of discrete-time nonlinear stochastic control systems. We impose conditions on the dt-SCS $\Sigma$ enabling us to find an SPSF from its infinite abstraction $\widehat{\Sigma}$ to $\Sigma$. The required conditions are presented via matrix inequalities.
\subsection{A Class of Nonlinear Stochastic Systems}\label{subsec:nonlinear}
Here, we focus on a specific class of discrete-time nonlinear stochastic control systems $\Sigma$ and quadratic stochastic pseudo-simulation functions $S$ and provide an approach on the construction of their infinite abstractions. The class of nonlinear systems is given by
\begin{align}\label{Eq_58a}
\Sigma:\left\{\hspace{-1.5mm}\begin{array}{l}x(k+1)=Ax(k)+E\varphi(Fx(k))+B\nu(k)+Dw(k)+R\varsigma(k),\\
y(k)=Cx(k),\end{array}\right.
\end{align}
where the additive noise $\varsigma(k)$ is a sequence of independent random vectors with multivariate standard normal distributions, and $\varphi:\R\rightarrow\R$ satisfies 
\begin{equation}\label{Eq_6a}
a\leq\frac{\varphi(c)-\varphi(d)}{c-d}\leq b,\quad\forall c,d\in\R,c\neq d,
\end{equation}
for some $a\in\R$ and $b\in\R_{>0}\cup\{\infty\}$, $a\leq b$. 

We use the tuple
\begin{align}\notag
\Sigma=(A,B,C,D,E,F,R,\varphi),
\end{align}
to refer to the class of nonlinear systems of the form~\eqref{Eq_58a}.
\begin{remark}
	If $E$ is a zero matrix or $\varphi$ in~\eqref{Eq_58a} is linear including the zero function (i.e., $\varphi\equiv0$), one can remove or push the term $E\varphi(Fx)$ to $Ax$, and consequently the nonlinear tuple reduces to the linear one $\Sigma=(A,B,C,D,R)$. Then, every time we mention the tuple $\Sigma=(A,B,C,D,E,F,R,\varphi)$, it implicitly implies that $\varphi$ is nonlinear and $E$ is nonzero. 
\end{remark}

\begin{remark}
	Without loss of generality and as mentioned in~\cite{arcak2001observer}, we can assume $a=0$ in~\eqref{Eq_6a} for the class of nonlinear control systems in~\eqref{Eq_58a}. If $a\neq0$, one can define a new function $\tilde\varphi(s):=\varphi(s)-as$ satisfying~\eqref{Eq_6a} with $\tilde a=0$ and $\tilde b=b-a$, and rewrite~\eqref{Eq_58a} as
	\begin{align}\notag
	\Sigma:\left\{\hspace{-1.5mm}\begin{array}{l}x(k+1)=\tilde Ax(k)+E\tilde \varphi(Fx(k))+B\nu(k)+Dw(k)+R\varsigma(k),\\
	y(k)=Cx(k)\end{array}\right.
	\end{align}
	where $\tilde A=A+aEF$.
\end{remark}

\begin{remark}\label{Re_1a}	
	We restrict ourselves here to systems with a single nonlinearity as in~\eqref{Eq_58a} for the sake of the simple presentation. However, it would be straightforward to show similar results for systems with multiple nonlinearities as
	\begin{align}\notag
	\Sigma:\left\{\hspace{-1.5mm}\begin{array}{l}x(k+1)=Ax(k)+\sum_{i=1}^{\bar M}E_i \varphi_i(F_ix(k))+B\nu(k)+Dw(k)+R\varsigma(k),\\
	y(k)=Cx(k),\end{array}\right.
	\end{align}
	where $\varphi_i:\R\rightarrow\R$ satisfies~\eqref{Eq_6a} for some $a_i\in\R$ and $b_i\in\R_{>0}\cup\{\infty\}$, for any $i\in\{1,\ldots,\bar M\}$.
\end{remark}

Here, we employ a quadratic SPSF of the form
\begin{align}\label{Eq_77a}
S(x,\hat x)=(x-P\hat x)^T M(x-P\hat x),
\end{align}
where $P$ and $ M\succ0$ are matrices of appropriate dimensions. In order to show that $S$ in~\eqref{Eq_77a} is an SPSF from $\widehat \Sigma$ to $\Sigma$, we require the following key assumption on $\Sigma$. 
\begin{assumption}\label{As_11a}
	Assume that for some constant
	$0<\hat\kappa<1$, there exist matrices $ M\succ0$, $K$, and $L_1$ of appropriate dimensions such that the matrix inequality~\eqref{Eq_8a} holds. Note that the left-hand side matrix in \eqref{Eq_8a} is symmetric as well. 
	\begin{figure*}
		\begin{align}\label{Eq_8a}
		\begin{bmatrix}
		(1+2/\pi)(A+BK)^T M(A+BK) && (A+BK)^T M(BL_1+E)\\
		*&& (1+2/\pi)(B\tilde R-P\hat B)^T  M(B\tilde R-P\hat B)
		\end{bmatrix}
		\preceq\begin{bmatrix}
		\hat\kappa M& -F^T\\
		-F & \frac{2}{b}
		\end{bmatrix}
		\end{align}
		\rule{\textwidth}{0.1pt}
	\end{figure*}
\end{assumption}

Now, we provide one of the main results of this section showing conditions under which $S$ in~\eqref{Eq_77a} is an SPSF from $\widehat \Sigma$ to $\Sigma$.
\begin{theorem}\label{Thm_33a}
	Let $\Sigma$
	and $\widehat \Sigma$ be two stochastic control subsystems. Suppose Assumption~\ref{As_11a} holds and there exist matrices $P$, $Q$, $S$, and $L_2$ such that
	\begin{IEEEeqnarray}{rCl}\IEEEyesnumber\label{Con_1056}
		\IEEEyessubnumber\label{Eq_10a} AP&=&P\hat A-BQ,\\
		\IEEEyessubnumber\label{Eq_15a}  E&=&P\hat E-B(L_1-L_2),\\
		\IEEEyessubnumber\label{Eq_14a}  D&=&P\hat D-\hat BS,\\
		\IEEEyessubnumber\label{Eq_11a}  R&=&P\hat R,\\
		\IEEEyessubnumber\label{Eq_16a}  \hat F&=&FP,\\
		\IEEEyessubnumber\label{Eq_26a}  \hat C&=& CP.
	\end{IEEEeqnarray}	
	Then function $S$ defined in~\eqref{Eq_77a} is an SPSF from $\widehat \Sigma$ to $\Sigma$.
\end{theorem}
The proof of Theorem~\ref{Thm_33a} is provided in Appendix.
Note that functions $\alpha,\kappa\in\mathcal{K}_\infty$, and $\rho_{\mathrm{int}}$, $\rho_{\mathrm{ext}}\in\mathcal{K}_\infty\cup\{0\}$ in Definition~\ref{Def_1a} associated with $S$ in~\eqref{Eq_77a} are defined as $\alpha(s)=\frac{\lambda_{\min}(M)}{
	n\lambda_{\max}(C^TC)}\,s^2$, $\kappa(s):=(1-(1-\tilde \pi)\tilde \kappa)\,s$, $\rho_{\mathrm{int}}(s):=(1+\tilde \delta) (\frac{1}{\tilde \kappa \tilde\pi})(p(1+2\pi+1/\pi))\Vert\sqrt{ M}D\Vert_2^2\,s^2$, $\rho_{\mathrm{ext}}(s):=(1+1/\tilde \delta)(\frac{1}{\tilde \kappa \tilde\pi})(m(1+3\pi)\Vert\sqrt{M}(B\tilde R-P\hat B)\Vert_2^2\, s^2$, $\forall s\in\mathbb R_{\ge0}$ where $\tilde \kappa = 1- \hat\kappa$, and constants $0<\tilde \pi<1$ and $\tilde \delta> 0$ can be chosen arbitrarily. Moreover, the positive constant $\psi$ in~\eqref{Eq_3a} is equal to zero here.

\begin{remark}
	Note that for any linear system $\Sigma=(A,B,C,D, R)$, stabilizability of the pair~$(A,B)$ is sufficient to satisfy Assumption~\ref{As_11a} in where matrices $E$, $F$, and $L_1$ are identically zero.
\end{remark}

\begin{remark}
	Since the results in Theorem~\ref{Thm_33a} do not impose any condition on the matrix $\hat B$, it can be arbitrarily chosen. One can choose $\hat B=\mathds{I}_{\hat n}$ to construct a fully actuated  infinite abstract system $\widehat \Sigma$, and consequently make the synthesis problem over it much easier.
\end{remark}

\begin{remark}
	Since Theorem~\ref{Thm_33a} does not impose any condition also on the matrix $\tilde R$, one can choose $\tilde R$ such that it minimizes the function $\rho_{\mathrm{ext}}$ which is given by \cite{girard2009hierarchical}:
	\begin{align}\notag
	\tilde R=(B^T M B)^{-1} B^T M P\hat B.
	\end{align}
\end{remark}

In the next section, we present a computational scheme to construct finite MDPs together with their corresponding stochastic pseudo-simulation functions for concrete models or their reduced-order versions. Note that we provide compositional frameworks for infinite and finite abstractions separately since one may be interested in employing one of the proposed results. In addition, if construction of infinite abstractions provided in Section~\ref{sec:constrcution_infinite} is not possible for some given dynamics, one can readily utilize the proposed results for finite abstractions (without performing the model order reduction) which is always possible as in the next section.

\section{Compositional Finite Abstractions for Interconnected Systems}\label{sec:compositionality1}

In this section, we consider $\Sigma_i=(X_i, U_i, W_i,\varsigma_i, f_i, Y_{i}, h_{i})$ as the original subsystems (or their reduced-order versions constructed in the previous section) and $\widehat \Sigma_i$ as their finite abstractions given by the tuple
\begin{equation*}
\widehat\Sigma_i=(\hat X_i,\hat U_i,\hat W_i,\varsigma_i,\hat f_i,\hat Y_{i},\hat h_{i}),	
\end{equation*}
with the input-output configuration similar to \eqref{config1} and \eqref{config2}, where $\hat W_i\subseteq  W_{i}$ and $\hat Y_i\subseteq  Y_{i}$. Moreover, we assume there exists an SPSF $S_i$ from $\widehat\Sigma_i$ to $\Sigma_{i}$ with the corresponding functions and constants denoted by $\alpha_i$, $\kappa_i$, $\rho_{\text{int}i}$, $\rho_{\text{ext}i}$, and $\psi_i$. In order to provide another compositionality result of the paper for interconnected \emph{finite} systems, we first define an abstraction map $\Pi_{w_{ji}}$ on $W_{ji}$ that assigns to any $w_{ji}\in W_{ji}$ a representative point $\bar w_{ji}\in\hat W_{ji}$ of the corresponding partition set containing $w_{ji}$. The mentioned map satisfies 
\begin{equation}
\label{eq:Pi_mu}
\Vert \Pi_{w_{ji}}(w_{ji})-w_{ji} \Vert \leq \mu_{ji},\,\quad \forall w_{ji}\in W_{ji},
\end{equation}	
where $\mu_{ji}$ is an \emph{internal input} discretization parameter defined similar to $\delta$ later in~\eqref{eq:Pi_delta}. Now we define a notion of the interconnection applicable to finite MDPs.
\begin{definition}
	Consider $N\in\mathbb N_{\geq1}$ finite stochastic control subsystems $\widehat \Sigma_i=(\hat X_i,\hat U_i,\hat W_i,\varsigma_i,\hat f_i,\hat Y_{i},\hat h_{i})$, $i\in \{1,\dots,N\}$. The interconnection of  $\widehat \Sigma_i$ is the \emph{finite} interconnected stochastic control system $\widehat \Sigma=(\hat X,\hat U,\varsigma,\hat f,\hat Y,\hat h)$, denoted by
	$\widehat {\mathcal{I}}(\widehat\Sigma_1,\ldots,\widehat\Sigma_N)$, such that $\hat X:=\prod_{i=1}^{N}\hat X_i$,  $\hat U:=\prod_{i=1}^{N}\hat U_i$, $\hat f:=\prod_{i=1}^{N}\hat f_{i}$, $\hat Y:=\prod_{i=1}^{N}\hat Y_{ii}$, and $\hat h=\prod_{i=1}^{N}\hat h_{ii}$, subjected to the following constraint:
	\begin{align}\notag
	\forall i,j\in \{1,\dots,N\},i\neq j\!: \quad \hat w_{ji} = \Pi_{w_{ji}}(\hat y_{ij}), \Pi_{w_{ji}}(\hat Y_{ij})\subseteq\hat W_{ji}.
	\end{align}
\end{definition}
Now we raise the following small-gain assumption similar to Assumption~\ref{Assump: Gamma}.

\begin{assumption}\label{Assump: Gamma1}
	Assume that there exist $\mathcal{K}_\infty$ functions $\tilde \delta_f, \bar \lambda$ such that $(\bar \lambda - \mathcal{I}_d)\in\mathcal{K}_\infty$ and $\mathcal{K}_\infty$ functions $\kappa_{ij}$ defined as
	\begin{equation*}
	\kappa_{ij}(s) := 
	\begin{cases}
	\kappa_i(s), \quad\quad& \text{if }i = j,\\
	(\mathcal{I}_d + \tilde \delta_f)\circ\rho_{\mathrm{int}i}\circ \bar \lambda \circ\alpha_j^{-1}(s), \quad\quad& \text{if }i \neq j,
	\end{cases}
	\end{equation*}
	satisfy
	\begin{equation}
	\label{Assump: Kappa1}
	\kappa_{i_1i_2}\circ\kappa_{i_2i_3}\circ\dots \circ \kappa_{i_{r-1}i_{r}}\circ\kappa_{i_{r}i_1} < \mathcal{I}_d
	\end{equation}	
	for all sequences $(i_1,\dots,i_{r}) \in \{1,\dots,N\}^{r}$ and ${r}\in \{1,\dots,N\}$.
	
	Similar to~\eqref{compositionality}, the small-gain condition~\eqref{Assump: Kappa1} implies the existence of $\mathcal{K}_\infty$ functions $\sigma_i>0$ \cite[Theorem 5.5]{ruffer2010monotone}, satisfying
	\begin{align}\label{compositionality1}
	\max_{i,j}\Big\{\sigma_i^{-1}\circ\kappa_{ij}\circ\sigma_j\Big\} < \mathcal{I}_d, \quad i,j = \{1,\dots,N\}.
	\end{align}
\end{assumption}	
In the next theorem, we leverage small-gain Assumption \ref{Assump: Gamma1} together with the concavity assumption of $\max_{i}\sigma_i^{-1}$ to quantify the error between the interconnection of stochastic control subsystems and that of their \emph{finite} abstractions in a compositional manner.

\begin{theorem}\label{Thm: Comp1}
	Consider the interconnected dt-SCS
	$\Sigma=\mathcal{I}(\Sigma_1,\ldots,\Sigma_{N})$ induced by $N\in\mathbb N_{\geq1}$ stochastic
	control subsystems~$\Sigma_{i}$. Suppose that each $\Sigma_{i}$ admits a finite abstraction $\widehat \Sigma_i$ together with an SPSF $S_i$. If Assumption~\ref{Assump: Gamma1} holds and $\max_{i}\sigma_i^{-1}$ for $\sigma_i$ as in \eqref{compositionality1} is concave,
	then the function $V(x,\hat x)$ defined as
	\begin{equation}
	\label{Comp: Simulation Function1}
	V(x,\hat x) := \max_{i} \Big\{\sigma_i^{-1}(S_i(x_{i},\hat x_i))\Big\},
	\end{equation}
	is an SSF from $\widehat \Sigma=\widehat {\mathcal{I}}(\widehat \Sigma_1,\ldots,\widehat\Sigma_N)$ to 	$\Sigma=\mathcal{I}(\Sigma_{1},\ldots,\Sigma_{N})$. 	
\end{theorem}

The proof of Theorem~\ref{Thm: Comp1} is provided in Appendix. Figure \ref{Fig1} shows schematically the results of Theorem \ref{Thm: Comp1}.

Next, we show how to construct finite Markov decision processes from concrete models (or their reduced-order versions).

\begin{figure}[ht]
	\begin{center}
		\includegraphics[width=11cm]{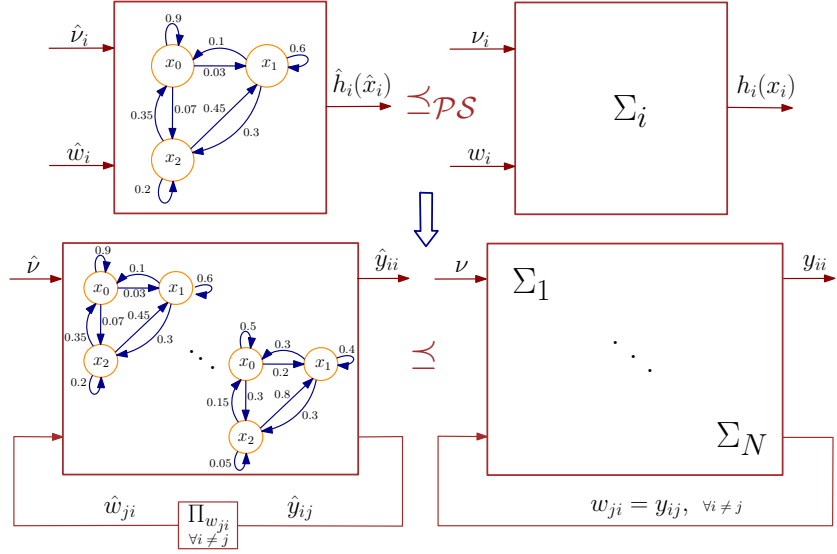}
		\caption{Compositionality results for constructing interconnected finite systems provided that the condition \eqref{Assump: Kappa1} is satisfied.}
		\label{Fig1}
	\end{center}
\end{figure}

\subsection{Finite Abstractions of dt-SCS}
\label{subsec:MDP}
Given a dt-SCS $\Sigma$, we construct its finite MDP $\widehat\Sigma$ as a finite abstraction of the original system. The abstraction algorithm works based on selecting finite partitions of state and input sets as $$X = \cup_i \mathsf X_i, ~U= \cup_i \mathsf U_i, ~W = \cup_i \mathsf W_i$$
and selection of representative points $\bar x_i\in \mathsf X_i$, $\bar \nu_i\in \mathsf U_i$, and $\bar w_i\in \mathsf W_i$ as abstract states and inputs.

Given a dt-SCS $\Sigma$, its finite abstract $\widehat\Sigma$ can be represented as 		
\begin{equation}
\label{eq:abs_tuple}
\widehat\Sigma =(\hat X, \hat U, \hat W, \varsigma,\hat f,\hat Y,\hat h),
\end{equation}
where $\hat X = \{\bar x_i,i=1,\ldots,n_x\}, \hat U = \{\bar u_i,i=1,\ldots,n_u\}$, $\hat W = \{\bar w_i,i=1,\ldots,n_w\}$ are the sets of selected representative points.
The function $\hat f:\hat X\times\hat U\times \hat W\times V_\varsigma\rightarrow\hat X$ is defined as
\begin{equation}
\label{eq:abs_dyn}
\hat f(\hat{x},\hat{\nu},\hat{w},\varsigma) = \Pi_x(f(\hat{x},\hat{\nu},\hat{w},\varsigma)),	
\end{equation}
where $\Pi_x:X\rightarrow \hat X$ is the map that assigns to any $x\in X$, the representative point $\bar x\in\hat X$ of the corresponding partition set containing $x$. The output map $\hat h$ is the same as $h$ with its domain restricted to the finite state set $\hat X$ and the output set $\hat Y$ is just the image of $\hat X$ under $h$. The initial state of $\widehat\Sigma$ is also selected according to $\hat x_0 := \Pi_x(x(0))$ with $ x(0)$ being the initial state of $\Sigma$.

We assume the abstraction map $\Pi_x$ used in \eqref{eq:abs_dyn} satisfies the inequality
\begin{equation}\label{eq:Pi_delta}
\Vert \Pi_x(x)-x\Vert \leq \delta,~\quad \forall x\,\in X,
\end{equation}
where $\delta$ is the \emph{state} discretization parameter defined as $\delta:=\sup\{\|x-x'\|,\,\, x,x'\in \mathsf X_i,\,i=1,2,\ldots,n_x\}$.

\begin{remark}
	Note that we do not have any requirements for discretizing the state, external, and internal input sets. However, the size of the state discretization parameter $\delta$ appears in the formulated error in~\eqref{general setting},~\eqref{Eq_305a}: one can decrease the error by reducing the state discretization parameter. We also do not have any constraints on the shape of partition elements in constructing finite MDPs. For the sake of an easy implementation, one can consider partition sets as boxes and the center of each box as representative points.
\end{remark}
\subsection{General Setting of Nonlinear Stochastic Systems}\label{subsec:general setting110}
In this subsection, we assume that the output map $h$ satisfies the following general Lipschitz assumption: there exists an $\tilde{\alpha}\in \mathcal{K}_{\infty}$ such that $\Vert h(x)-h(x')\Vert \leq \tilde{\alpha}(\Vert x-x'\Vert)$ for all $x,x' \in X$. Note that this assumption on $h$ is not restrictive provided that $h$ is continuous and one works on a compact subset of $X$. We impose conditions on the infinite dt-SCS $\Sigma$ enabling us to find SPSF from its finite abstraction $\widehat{\Sigma}$, constructed as in the previous subsection, to $\Sigma$. The existence of an SPSF is established under the assumption that the original model (or its reduced-order version) is \emph{incrementally input-to-state stable} as in the next definition.

\begin{definition}\label{Def111}
	A dt-SCS  $\Sigma$ is called \emph{incrementally input-to-state stable} if there exists a function $S: X\times X\to \mathbb{R}_{\geq0}$  such that $\forall x, x'\in X$, $\forall\nu,\nu'\in U$, $\forall w, w' \in W$, the following two inequalities
	
	\begin{align}\label{Con555}
	\underline{\alpha}(\Vert x-x'\Vert ) \leq S(x,x')\leq \overline{\alpha} (\Vert x-x'\Vert ),
	\end{align}	
	and	
	\begin{align}\notag
	\mathbb{E}& \Big[S(f(x,\nu,w,\varsigma),f(x',\nu',w',\varsigma))\big|x,x',\nu,\nu', w, w'\Big]-S(x,x')\\\label{Con854}
	&\leq-\bar{\kappa}(S(x,x'))+\bar \rho_{\mathrm{int}}(\Vert w-w'\Vert)+\bar \rho_{\mathrm{ext}}(\Vert\nu-\nu'\Vert),
	\end{align}
	hold for some $\underline{\alpha}, \overline{\alpha}, \bar{\kappa}\in \mathcal{K}_{\infty}$, and $\bar \rho_{\mathrm{int}}$, $\bar \rho_{\mathrm{ext}}\in\mathcal{K}_\infty\cup\{0\}$.
\end{definition}

\begin{remark}
	Note that the above definition is a stochastic counterpart of the incremental ISS Lyapunov functions defined for discrete-time deterministic systems~\cite{tran2016convergence}.
\end{remark}

In the next subsection, we show that inequalities \eqref{Con555}-\eqref{Con854} for a candidate quadratic function $S$ and a class of nonlinear stochastic control systems boil down to some matrix inequalities.

Now we show that under a mild condition, the function $S$, as in Definition~\ref{Def111}, is indeed an SPSF from $\widehat \Sigma$ to $\Sigma$.
\begin{theorem}\label{Thm_5a}
	Let $\Sigma$ be an incrementally input-to-state stable dt-SCS  via a function $S$ as in Definition~\ref{Def111} and $\widehat{\Sigma}$ be its finite MDP as in Subsection~\ref{subsec:MDP}. Assume that there exists a function $\gamma\in\mathcal{K}_{\infty}$  such that $S$ satisfies
	\begin{equation}\label{Eq65}
	S(x,x')-S(x,x'')\leq \gamma(\Vert x'-x''\Vert),\quad \forall x,x',x'' \in X.
	\end{equation}
	Then $S$ is a stochastic pseudo-simulation function from $\widehat{\Sigma}$ to $\Sigma$.
\end{theorem}
The proof of Theorem~\ref{Thm_5a} is provided in Appendix.

\begin{remark}
	Note that by employing the mean value theorem as in \cite{zamani2014symbolic},  the condition~\eqref{Eq65} is always satisfied for any differentiable function $S$ restricted to a compact subset of $X \times X$.
\end{remark}

Now we provide similar results as in this subsection but tailored to the particular class of nonlinear stochastic control systems.

\subsection{The Class of Nonlinear Stochastic Systems}\label{subsec:nonlinear110}
\begin{figure*}
	\begin{align}\label{Eq_88a}
	\begin{bmatrix}
	(1+2/\pi)(A+BK)^T M(A+BK) && (A+BK)^T ME\\
	*&& (1+2/\pi)E^T  ME
	\end{bmatrix}
	\preceq\begin{bmatrix}
	\hat\kappa M& -F^T\\
	-F & \frac{2}{b}
	\end{bmatrix}
	\end{align}
	\rule{\textwidth}{0.1pt}
\end{figure*}

Here, we focus on $\Sigma$ in~\eqref{Eq_58a} and propose an approach to construct its finite abstraction $\widehat \Sigma$. 
We candidate the following pseudo-simulation function
\begin{equation}
\label{Eq_7a}
S(x,\hat x)=(x-\hat x)^T M(x-\hat x),
\end{equation}
where $ M$ is a positive-definite matrix of an appropriate dimension. In order to show that $S$ in~\eqref{Eq_7a} is an SPSF from $\widehat\Sigma$ to $\Sigma$, we require the following assumption on $\Sigma$. 

\begin{assumption}\label{As_31a}
	Assume that for a constant
	$0<\hat\kappa<1$, there exist matrices $ M\succ0$, and $K$ of appropriate dimensions such that the inequality~\eqref{Eq_88a} holds. Note that the matrix in the left-hand side of the inequality \eqref{Eq_88a} is symmetric as well.
\end{assumption}
Now we provide another main result of this paper showing under which conditions $S$ in~\eqref{Eq_7a} is an SPSF from $\widehat \Sigma$ to $\Sigma$.
\begin{theorem}\label{Thm_3a}
	Assume the system $\Sigma$ satisfies Assumption~\ref{As_31a}. Let $\widehat \Sigma$ be its finite abstraction with the state discretization parameter $\delta$. Then the function $S$ defined in~\eqref{Eq_7a} is an SPSF from $\widehat \Sigma$ to $\Sigma$.
\end{theorem}
The proof of Theorem~\ref{Thm_3a} is provided in Appendix.

Note that the functions $\alpha,\kappa\in\mathcal{K}_\infty$, and $\rho_{\mathrm{int}}$, $\rho_{\mathrm{ext}}\in\mathcal{K}_\infty\cup\{0\}$ in Definition~\ref{Def_1a} associated with $S$ in~\eqref{Eq_7a} are defined as $\alpha(s)=\frac{\lambda_{\min}( M)}{n\lambda_{\max}(C^TC)}\,s^2$, $\kappa(s):=(1-(1-\tilde \pi)\tilde \kappa)\,s$, $\rho_{\mathrm{int}}(s):=(1+\tilde \delta) (\frac{1}{\tilde \kappa \tilde\pi})(p(1+2\pi+1/\pi))\Vert\sqrt{M}D\Vert_2^2\,s^2$, $\rho_{\mathrm{ext}}(s):=0$, $\forall s\in\mathbb R_{\ge0}$ where $\tilde \kappa = 1- \hat\kappa$ and constants $0<\tilde \pi <1$ and $\tilde \delta > 0$ can be chosen arbitrarily. Moreover, the positive constant $\psi$ in~\eqref{Eq_3a} is $\psi=(1+1/\tilde \delta)(\frac{1}{\tilde \kappa \tilde\pi})(n(1+3\pi)\lambda_{\max}{( M))}\,\delta^2$.

\section{Case Studies}\label{example}
\begin{figure}
	\centering
	\includegraphics[width=5cm]{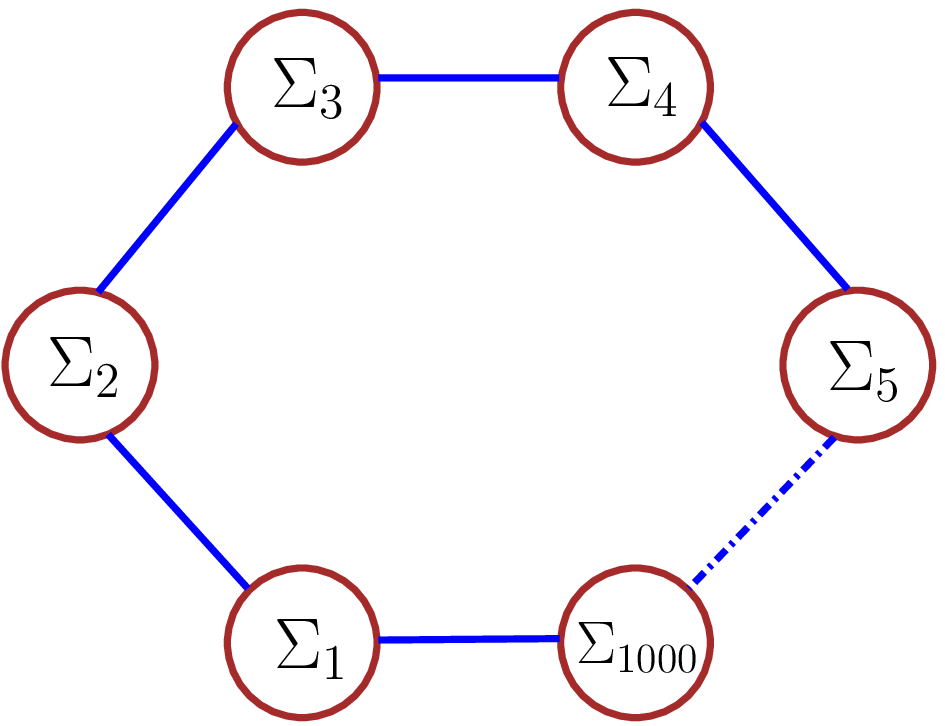}\hspace{1cm}
	\includegraphics[width=5cm]{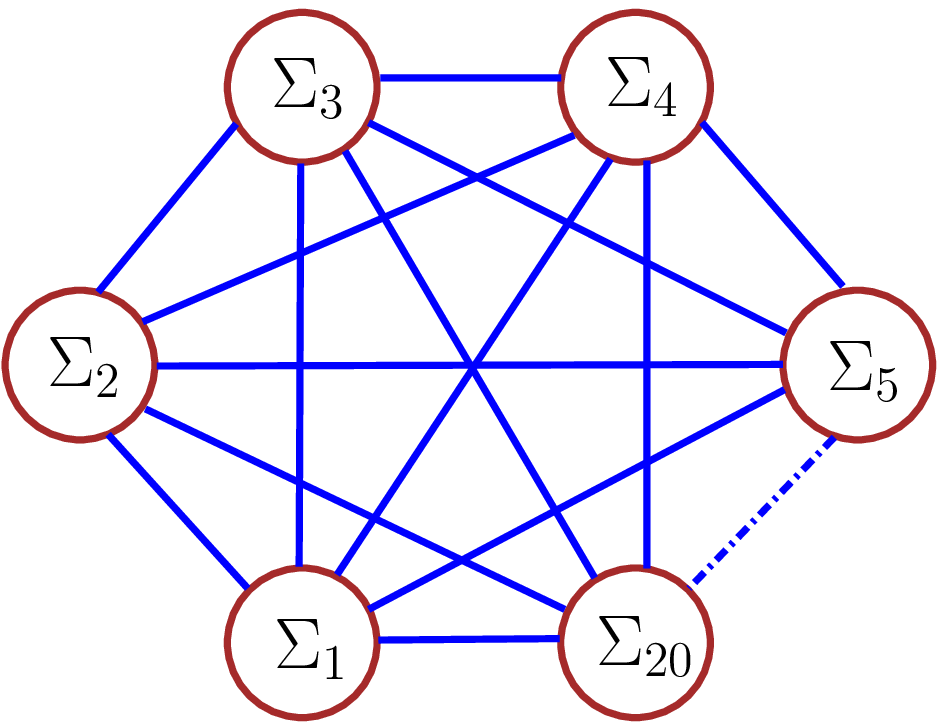}
	\caption{Left: A circular building in a network of $1000$ rooms. Right: A fully interconnected network of $20$ nonlinear components (totally $100$ dimensions).}
	\label{Case_Studies}
\end{figure}

We first apply our technique to a fully interconnected network of $20$ nonlinear subsystems (totally $100$ dimensions) and construct finite MDPs from their reduced-order versions (together $20$ dimensions) with guaranteed probabilistic error bounds on their output trajectories (cf. Figure~\ref{Case_Studies} right). We then apply our proposed approaches to a temperature regulation in a circular building (cf. Figure~\ref{Case_Studies} left) and construct compositionally a finite abstraction of the network containing $1000$ rooms. We employ the constructed finite abstractions as substitutes to compositionally synthesize policies regulating the temperature in each room for a bounded time horizon.

\subsection{Fully Interconnected Network}
In order to show the applicability of our approach to strongly connected networks with  nonlinear dynamics, we consider nonlinear dt-SCS
\begin{equation*}
\Sigma:\left\{\hspace{-1.5mm}\begin{array}{l}{x}(k+1)= Gx(k)+\varphi(x(k))+\nu(k)+R\varsigma(k),\\
y(k)=x(k),\end{array}\right.
\end{equation*}
for some matrix $G=(\mathds{I}_n-\tau L)\in \mathbb R^{n\times n}$ where $\tau L$ is the Laplacian matrix of an undirected graph with $0<\tau <1/\Delta$, and $\Delta$ is the maximum degree of the graph \cite{godsil2001}. 
We assume $L$ is the Laplacian matrix of a complete graph as
\begin{align}\label{Eq_90}
L=\begin{bmatrix}n-1 & -1 & \cdots & \cdots & -1 \\  -1 & n-1 & -1 & \cdots & -1 \\ -1 & -1 & n-1 & \cdots & -1 \\ \vdots &  & \ddots & \ddots & \vdots \\ -1 & \cdots & \cdots & -1 & n-1\end{bmatrix}_{n\times n}\!\!\!\!\!\!\!\!\!\!\!\!,
\end{align}
and $\tau = 0.001$. Moreover, $R = \mathsf{diag}(\mathds{1}_{n_1},\ldots,\mathds{1}_{n_N})$, $\varsigma(k)=[\varsigma_1(k);\ldots;\varsigma_N(k)]$, $\varphi(x)=[\mathds{1}_{n_1}\varphi_1(F_1 x_1(k));\ldots;\mathds{1}_{n_N}\varphi_N\\(F_N x_N(k))]$ where $n=\sum_{i=1}^Nn_i$, $\varphi_i(x) = sin(x)$, and $F_i^T= \begin{bmatrix}0.1& 0& \cdots & 0\end{bmatrix}^T\in\R^{n_i}$ $\forall i\in\{1,\ldots,N\}$. 
We partition $x(k)$ as $x(k)=[x_1(k);\ldots;x_N(k)]$ and $\nu(k)$ as $\nu(k)=[\nu_1(k);\ldots;\nu_N(k)]$, where $x_i(k),\nu_i(k)\in\R^{n_i}$. Now, we introduce $\Sigma_i$ as
\begin{equation*}
\Sigma_i:\left\{\hspace{-1.5mm}\begin{array}{l}x_i(k+1)=A_ix_i(k)+\mathds{1}_{n_i}\varphi_i(F_i x_i(k))+\nu_i(k)+D_i w_i(k)+\mathds{1}_{n_i}\varsigma_i(k),\\
y_i(k)=x_i(k),\\
\end{array}\right.
\end{equation*}
where $ A_{i} = (\mathds{I}_{n_i}-\tau L_{i})$, $w_i(k) = [{y_{1i};\ldots;y_{(i-1)i};y_{(i+1)i};\ldots;y_{Ni}}]$, $i\in \{1,\ldots,N\}$, and
\begin{align}\notag
&L_{i} =  \begin{bmatrix}n-1 &-1 & \cdots & -1 \\  -1& n-1 &\cdots&-1\\ \vdots & & \ddots&\vdots\\ -1 & \cdots & -1 & n-1\end{bmatrix}_{n_i\times n_i}\!\!\!\!\!\!\!\!\!\!\!\!\!\!\!\!,\\\notag
&D_i  =  -\tau\begin{bmatrix}-1 &-1 & \cdots & -1 \\  -1& -1 &\cdots&-1\\ \vdots & & \ddots&\vdots\\ -1 & \cdots & -1 & -1\end{bmatrix}_{n_i\times (n-n_i)}\!\!\!\!\!\!\!\!\!\!\!\!\!\!\!\!\!\!\!\!\!\!\!\!\!\!,~~~\forall i\in \{1,\dots,N\}.
\end{align}
We fix $N=20$, $n=100$, $n_i=5$, $\forall i\in\{1,\ldots,N\}$. Then one can readily verify that $\Sigma=\mathcal{I}(\Sigma_1,\ldots,\Sigma_N)$. Our goal is to first aggregate each $x_i$ into a scalar-valued $\hat x_{{\textsf r}i}$ (the index ${\textsf r}$ signifies the reduced-order version of the original model), governed by $\widehat \Sigma_{{\textsf r}i}$, which satisfies:
\begin{align}\notag
\widehat \Sigma_{{\textsf r}i}:\left\{\hspace{-1.5mm}\begin{array}{l}\hat x_{{\textsf r}i}(k+1)=0.5\hat x_{{\textsf r}i}(k)+0.1\varphi_i(0.1\hat x_{{\textsf r}i}(k))+\hat \nu_{{\textsf r}i}(k)+\hat D_i\hat w_{{\textsf r}i}(k)+\varsigma_i(k),\\
\hat y_{{\textsf r}i}(k)=\hat C_i\hat x_{{\textsf r}i}(k),\end{array}\right.
\end{align}
where $\hat D_i =0.001\mathds{1}_{95}^T$, $\hat C_i =\mathds{1}_{5}$, and $\hat w_{{\textsf r}i}(k) \in\R^{95}$. One can readily verify that, for any $i\in\{1,\ldots,N\}$, the condition~\eqref{Eq_8a} is satisfied with $ M_{i}=\mathds{I}_{5}$, $\hat \kappa_{i}=0.003$, $\pi_{i}=1$, $P_{i}=\mathds{1}_{5}$, $L_{1i}=-\mathds{1}_{5}$, $\tilde R_i =\mathds{1}_{5}$, $b_{i} = 1$, and $K_{i}$ as a $5\times 5$ matrix with diagonal elements $-0.9$, and off-diagonals $-0.001$. Moreover, for any $i\in\{1,\ldots,N\}$, conditions~\eqref{Con_1056} are satisfied by ${L_2}_i=-0.1\mathds{1}_{5}$, $Q_i = -0.4\mathds{1}_{5}$, and $S_i = \mathbf{0}_{5\times 95}$.
We fix SPSF as in~\eqref{Eq_77a}. By taking $\tilde \pi_{i} = 0.99$, $\tilde \kappa_{i} = 0.99$ and $\tilde \delta_{i} = 0.1$, $\forall i\in\{1,\ldots,N\}$, one can verify that $S_{i}(x_i,\hat x_{{\textsf r}i})=(x_i-\mathds{1}_{5}\hat x_{{\textsf r}i})^T\mathds{I}_{5}(x_i-\mathds{1}_{5}\hat x_{{\textsf r}i})$ is an SPSF from $\widehat\Sigma_{{\textsf r}i}$ to $\Sigma_i$ satisfying the condition \eqref{Eq_2a} with $\alpha_{i}(s)=1/5s^2$ and the condition \eqref{Eq_3a} with $\kappa_{i}(s)=0.99s$, $\rho_{\mathrm{int}i}(s)=0.2s^2$, $\rho_{\mathrm{ext}i}(s)=0$, $\forall s\in \mathbb R_{\ge0}$, and $\psi_{i}=0$, where the input $\nu_i$ is given via the interface function in~\eqref{Eq_40a} as
\begin{align}\notag
\nu_i=&-K_{i}(x_i-\mathds{1}_{5}\hat x_{{\textsf r}i})- 0.4\mathds{1}_{5}\hat x_{{\textsf r}i} + \mathds{1}_{5}\hat \nu_{{\textsf r}i}- \mathds{1}_{5} \varphi_i (F_i x_i)+ 0.1\mathds{1}_{5} \varphi_i (F_i \mathds{1}_{5} \hat x_{{\textsf r}i}). 
\end{align} 
By taking $\sigma_{i}(s) = s$, $ \forall i\in\{1,\ldots,N\}$ , one can readily verify that the small-gain condition~\eqref{Assump: Kappa} and as a result the condition \eqref{compositionality} are satisfied. Hence, $V(x,\hat x_{\textsf r})=\max_{i} (x_i-\mathds{1}_{5}\hat x_{{\textsf r}i})^T\mathds{I}_{5}(x_i-\mathds{1}_{5}\hat x_{{\textsf r}i})$ is an SSF from  $\widehat\Sigma_{\textsf r}=\mathcal{I}( \widehat \Sigma_{{\textsf r}1},\ldots, \widehat \Sigma_{{\textsf r}N})$ to $\Sigma$ satisfying conditions \eqref{eq:lowerbound2} and \eqref{eq6666}  with $\alpha(s)=1/25s^2$, $\kappa(s)=0.99\,s$, $\rho_{\mathrm{ext}}(s)=0$, $\forall s\in \mathbb R_{\ge0}$, and $\psi=0$.

By starting the interconnected original system $\Sigma$ from $\mathbf{0}_{100}$ and its infinite abstraction $\widehat \Sigma_{\textsf r}$ from $\mathbf{0}_{20}$, using Theorem \ref{Thm_1a} and since $\psi=0$, we guarantee that the mismatch between outputs of $\Sigma$ and $\widehat \Sigma_{\textsf r}$ will not exceed $\varepsilon_1 = 0.25$ during the time horizon $T_d=100$ with the probability one.

Now we proceed with finding an SPSF from the finite MDP $\widehat\Sigma_i$ to the reduced-order model $\widehat\Sigma_{{\textsf r}i}$. One can readily verify that, for any $i\in\{1,\ldots,N\}$, the condition~\eqref{Eq_88a} is satisfied with $ M_{i}=1$, $\hat \kappa_{i}=0.009$, $\pi_{i}=1$, $K_{i}=-0.49$, and $b_{i} = 1$. By taking $\tilde \pi_{i} = 0.99$, $\tilde \kappa_{i} = 0.99$ and $\tilde \delta_{i} = 0.9$\, $\forall i\in\{1,\ldots,N\}$, the function $S_i(\hat x_{{\textsf r}i},\hat x_i)=(\hat x_{{\textsf r}i}-\hat x_i)^2$ is an SPSF from $\widehat\Sigma_{i}$ to $\widehat\Sigma_{{\textsf r}i}$ satisfying the condition \eqref{Eq_2a} with $\alpha_{i}(s)=1/5s^2$ and the condition \eqref{Eq_3a} with $\kappa_i(s)=0.99s$, $\rho_{\mathrm{int}i}(s)=0.26s^2$, $\rho_{\mathrm{ext}i}(s)=0$, $\forall s\in \mathbb R_{\ge0}$, and $\psi_i=8.42\delta^2$, where the input $\nu_i$ is given via the interface function in~\eqref{Eq_255} as
\begin{align}\notag
\hat \nu_{{\textsf r}i}=&-0.49(\hat x_{{\textsf r}i}-\hat x_i)+ \hat \nu_{i}. 
\end{align} 

By taking $\sigma_{i}(s) = s$, $ \forall i\in\{1,\ldots,N\}$ , one can readily verify that the small-gain condition~\eqref{Assump: Kappa1} and as a result condition \eqref{compositionality1} are satisfied. Hence, $V(\hat x_{\textsf r},\hat x)=\max_{i} (\hat x_{{\textsf r}i}-\hat x_i)^2$ is an SSF from  $\widehat\Sigma=\widehat{\mathcal{I}}(\widehat\Sigma_1,\ldots,\widehat\Sigma_N)$, with $\mu_{ji}=0$ $\forall i,j\in\{1,\ldots,N\}$, $i\neq j$, to $\widehat\Sigma_{\textsf r}$ satisfying conditions \eqref{eq:lowerbound2} and \eqref{eq6666}  with $\alpha(s)=1/25s^2$, $\kappa(s)=0.99\,s$, $\rho_{\mathrm{ext}}(s)=0$, $\forall s\in \mathbb R_{\ge0}$, and $\psi=8.42\delta^2$.

By taking the state discretization parameter $\delta = 0.001$, starting the interconnected infinite abstraction $\widehat \Sigma_{\textsf r}$ and its finite version $\widehat \Sigma$ from $\mathbf{0}_{20}$, and using Theorem \ref{Thm_1a}, we guarantee that the mismatch between outputs of $\widehat \Sigma_{\textsf r}$ and $\widehat \Sigma$ will not exceed $\varepsilon_2 = 0.25$ during the time horizon $T_d=100$ with the probability at least $92\%$.

Now we leverage Proposition~\ref{proposition} to provide a probabilistic closeness guarantee between the interconnected original system $\Sigma$ and the finite abstraction $\widehat \Sigma$. By taking $\delta = 0.001$, starting the interconnected systems $\Sigma$ from $\mathbf{0}_{100}$, $\widehat \Sigma_{\textsf r}$ and $ \widehat \Sigma$ from $\mathbf{0}_{20}$, and using Theorem \ref{Thm_1a} and Proposition~\ref{proposition}, we guarantee that the mismatch between outputs of $\Sigma$ and $\widehat \Sigma$ will not exceed $\varepsilon = 0.5,  (\varepsilon_1 = \varepsilon_2 = 0.25)$, during the time horizon $T_d=100$ with the probability at least $92\%$, , i.e.,
\begin{align}\notag
\mathbb P(\Vert y(k)-\hat y(k)\Vert\le 0.5,\,\, \forall k\in[0,100])\ge 0.92\,.
\end{align}

\begin{figure}
	\centering
	\includegraphics[width=10cm]{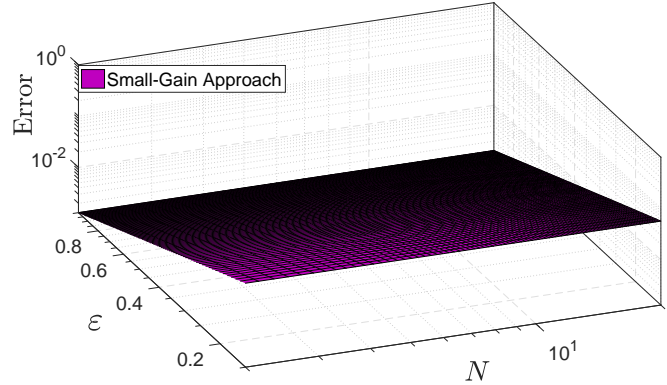}
	\caption{Fully interconnected network: Error bound in \eqref{Eq_25} provided by our approach based on small-gain conditions. Plot is in the logarithmic scale for a fixed $\delta = 0.001$ and $T_d = 100$. By increasing the number of subsystems, the error provided in~\eqref{Eq_25} does not change since the overall $\psi$ is independent of the size of the network (i.e., $N$), and is computed only based on the maximum $\psi_i$ of subsystems instead of being a linear combination of them which is the case in \cite{lavaei2017HSCC}.}
	\label{Laplacian}
\end{figure}

In Figure~\ref{Laplacian} which is in the logarithmic scale, we have fixed $\delta=0.001$ and plotted the error (the upper bound of the probability in \eqref{Eq_25}) as a function of the number of subsystems $N$ and the confidence bound $\varepsilon$ (cf. \eqref{Eq_25}). As seen, the overall $\psi$ in \eqref{Eq_25} is independent of the size of the network, and is computed based on the \emph{maximum} of $\psi_i$ of subsystems instead of being a \emph{linear combination} of them which is the case in \cite{lavaei2017HSCC}. Hence, by increasing the number of subsystems, the error does not change.

Note that one can follow the same compositional synthesis approach discussed above by constructing finite abstractions directly for subsystems without using infinite abstractions (model-order reductions). However, constructing finite abstractions for $5$-dimensional subsystems result in very large finite MDPs and may not be possible with the limited computational and memory resources.
The main benefit of infinite abstractions here is to help reducing dimensions of subsystems to scaler ones and then construct finite abstractions for the reduced-order versions while still providing the probabilistic closeness guarantee.

\subsection{Room Temperature Network}
Consider a network of $n\geq 3$ rooms each equipped with a heater and connected circularly (cf. Figure~\ref{Case_Studies} left).
The model of this case study is adapted from~\cite{meyer} by including stochasticity in the model as an additive noise. The evolution of temperatures $T$ 
can be described by the interconnected linear \mbox{dt-SCS}
\begin{equation*}
\Sigma:\left\{\hspace{-1.5mm}\begin{array}{l}{T}(k+1)=\bar A{T}(k)+\gamma T_{h}\nu(k)+ \beta T_{E}+\varsigma(k),\\
y(k)={T}(k),\end{array}\right.
\end{equation*}
where $\bar A$ is a matrix with diagonal elements $\bar a_{ii}=(1-2\eta-\beta-\gamma\nu_{i}(k))$, $i\in\{1,\ldots,n\}$, off-diagonal elements $\bar a_{i,i+1}=\bar a_{i+1,i}=\bar a_{1,n}=\bar a_{n,1}=\eta$, $i\in \{1,\ldots,n-1\}$, and all other elements are identically zero.
Parameters $\eta$, $\beta$, and $\gamma$ are conduction factors, respectively, between rooms $i \pm 1$ and the room $i$, between the external environment and the room $i$, and between the heater and the room $i$. Moreover,  $ T(k)=[T_1(k);\ldots;T_n(k)]$,  $\nu(k)=[\nu_1(k);\ldots;\nu_n(k)]$, $ \varsigma(k)=[\varsigma_1(k);\ldots;\varsigma_n(k)]$, $T_E=[T_{e1};\ldots;T_{en}]$, where $T_i(k)$ and $\nu_i(k)$ are taking values in sets $[19,21]$ and $[0,0.6]$, respectively, for all $i\in\{1,\ldots,n\}$. Outside temperatures are the same for all rooms: $T_{ei}=-1\,^\circ C$, $\forall i\in\{1,\ldots,n\}$, and the heater temperature $T_h=50\,^\circ C$.
Let us consider the individual rooms as $\Sigma_i$ described as
\begin{equation*}
\Sigma_i:\left\{\hspace{-1.5mm}\begin{array}{l}T_i(k+1)=A_iT_i(k)+\gamma T_{h} \nu_i(k)+D_i w_i(k)+\beta T_{ei}+\varsigma_i(k),\\
y_i(k)=T_i(k),\\
\end{array}\right.
\end{equation*}
where $A_i = \bar a_{ii},\, i\in \{1,\ldots,n\}$. One can readily verify that $\Sigma=\mathcal{I}(\Sigma_1,\ldots,\Sigma_N)$ where $D_i = [\eta;\eta]^T$, and $w_i(k) = [y_{i-1}(k);y_{i+1}(k)]$ (with $y_0 = y_n$ and $y_{n+1} = y_1$). Note that since the dynamic of each room is scaler (no need to reduce the order), our objective here is just to construct the finite abstraction of each room. First, we fix the SPSF as in~\eqref{Eq_7a}. Since the dynamic of the system is linear, the condition~\eqref{Eq_88a} reduces to 
\begin{align}\notag
(1+2/\pi_i)(A_{i}+B_{i}K_i)^T M_i(A_{i}+B_{i}K_i)\preceq  \hat\kappa_i  M_i, \label{Eq9a}
\end{align}
which is nothing more than the stabilizability of the temperature dynamic in room $i$. One can readily verify that this condition is satisfied with $ M_i=1$, $K_i=0$, $\pi_i = 1$, $\hat\kappa_i = 0.48$ $\forall i\in\{1,\ldots,n\}$, and $\eta = 0.1, \beta=0.4, \gamma = 0.5$. Then the function $S_i(T_i,\hat T_i)=(T_i-\hat T_i)^2$ is an SPSF from $\widehat\Sigma_i$ to $\Sigma_i$ satisfying the condition \eqref{Eq_2a} with $\alpha_{i}(s)=s^2$ and the condition \eqref{Eq_3a} with $\kappa_i(s)=0.99s$, $\rho_{\mathrm{int}i}(s)=0.91s^2$, $\rho_{\mathrm{ext}i}(s)=0$, $\forall s\in \mathbb R_{\ge0}$, and $\psi_i = 7.6\,\delta_i^2$.

Now we check the small-gain condition~\eqref{Assump: Kappa1} that is required for the compositionality result. By taking $\sigma_i(s) = s$, $\forall i\in\{1,\ldots,n\}$, the condition~\eqref{Assump: Kappa1} and as a result the condition \eqref{compositionality1} are always satisfied without any restriction on the number of rooms.
Hence, $V(T,\hat T)=\max_{i} (T_i-\hat T_i)^2$ is an SSF from  $\widehat\Sigma$ to $\Sigma$ satisfying conditions \eqref{eq:lowerbound2} and \eqref{eq6666}  with $\alpha(s)=s^2$, $\kappa(s)=0.99\,s$, $\rho_{\mathrm{ext}}(s)=0$, and $\psi = 7.6 \,\delta^2$.

We fix $n=1000$ and set the state discretization parameter $\delta = 0.005$. The initial states of the interconnected systems $\Sigma$ and $ \widehat \Sigma$ are selected as $20\mathds{1}_{1000}$. Using Theorem \ref{Thm_1a}, we guarantee that the distance between outputs of $\Sigma$ and $\widehat \Sigma$ will not exceed $\varepsilon = 0.5$ during the time horizon $T_d=100$ with the probability at least $98\%$, i.e.,
\begin{equation}
\label{eq:guarantee}
\mathbb P(\Vert y(k)-\hat y(k)\Vert\le 0.5,\,\, \forall k\in[0,100])\ge 0.98\,.
\end{equation}
Note that for the construction of finite abstractions, we have selected the center of partition sets as representative points. Moreover, we assume $\hat Y_{ij}= \hat W_{ji}$.

\begin{figure}
	\centering
	\includegraphics[width=9.2cm]{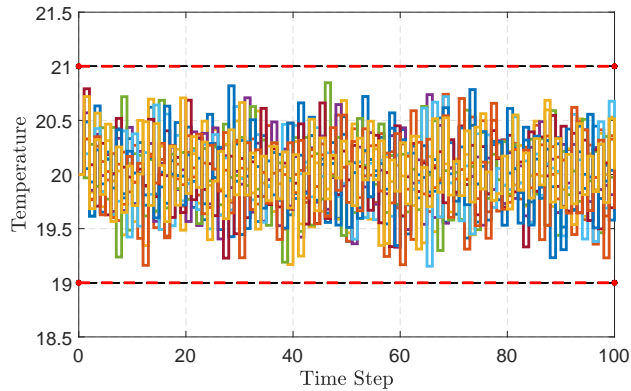}
	\caption{Closed-loop state trajectories of a representative room with different noise realizations in a network of $1000$ rooms.}
	\label{Simulation}
\end{figure}

Let us now synthesize a controller for $\Sigma$ via the abstraction $\widehat \Sigma$ such that the controller maintains the temperature of any room in the comfort zone $[19,21]$. We design a local controller for the abstract subsystem $\widehat \Sigma_i$, and then refine it back to the subsystem $\Sigma_i$ using the interface function. We employ the tool \software~\cite{FAUST15} to synthesize controllers for $\Sigma_i$ by taking the external input discretization parameter as $0.04$ and the standard deviation of the noise as $0.21$, $\forall i\in\{1,\ldots,n\}$. Closed-loop state trajectories of a representative room with different noise realizations are illustrated in Figure~\ref{Simulation} with only $10$ trajectories.

\begin{figure}
	\centering
	\includegraphics[width=10cm]{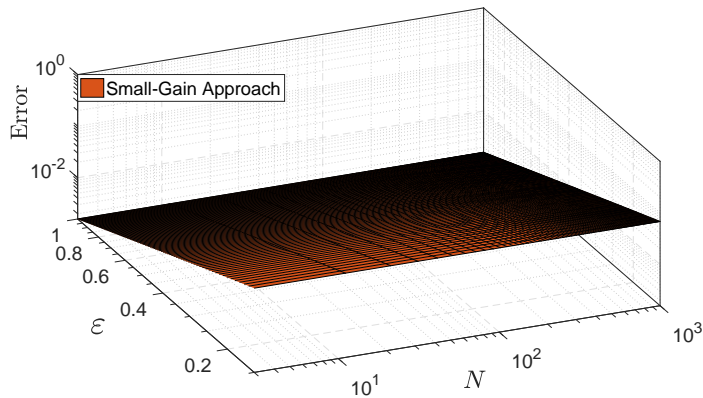}
	\caption{Room temperature network: Error bound in \eqref{Eq_25} provided by our approach based on small-gain conditions. Plot is in the logarithmic scale for a fixed $\delta = 0.005$, and $T_d = 100$. By increasing the number of subsystems, the error provided in~\eqref{Eq_25} does not change since the overall $\psi$ is independent of the size of the network (i.e., $N$), and is computed only based on the maximum $\psi_i$ of subsystems instead of being a linear combination of them which is the case in \cite{lavaei2017HSCC}.}
	\label{Temperature_Control}
\end{figure}

Similarly, we have fixed $\delta=0.005$ and plotted in Figure \ref{Temperature_Control} the error between the finite MPD $\widehat \Sigma$ and the concrete model $\Sigma$ as a function of the number of subsystems $N$ and the confidence bound $\varepsilon$. As seen, by increasing the number of subsystems, the error does not change since the overall $\psi$ in~\eqref{Eq_25} is independent of the size of the network.

\section{Discussion}
In this paper, we provided a compositional methodology for constructing both infinite abstractions (reduced order models) and finite MDPs of large-scale stochastic control systems. First, we introduced new notions of stochastic pseudo-simulation and simulation functions to compute the probabilistic mismatch between the concrete systems and their infinite abstractions. We then provided a compositional scheme on the construction of infinite abstractions of interconnected systems using the small-gain type reasoning. Accordingly, we constructed infinite abstractions together with their corresponding stochastic simulation functions for a particular class of nonlinear stochastic systems. Afterwards, we leveraged small-gain type conditions for the compositional construction of finite abstractions. We also proposed an approach to construct finite MDPs from concrete models (or their reduced-order versions) for stochastic control systems satisfying incremental input-to-state stability property. We showed that for the particular class of nonlinear systems, the aforementioned property can be readily checked by matrix inequalities. Finally, we applied our technique to a fully interconnected network and constructed finite MDPs from their reduced-order versions. We also applied our approach to the temperature regulation in a circular building and employed the constructed finite abstractions as substitutes to compositionally synthesize policies regulating the temperature in each room for a bounded time horizon. Compositional controller synthesis for large-scale stochastic systems is under investigation as a future work.

\bibliographystyle{alpha}
\bibliography{biblio}

\section{Appendix}
\begin{definition}\label{Def_Com1}
	Consider two dt-SCS 	$\Sigma =(X,U,\varsigma,f, Y,h)$ and
	$\widehat\Sigma =(\hat X,\hat U,\varsigma,\hat f, \hat Y,\hat h)$ without internal signals, where $\hat Y\subseteq Y$.
	A function $V:X\times\hat X\to\mathbb R_{\ge0}$ is
	called a stochastic simulation function (SSF) from $\widehat\Sigma$  to $\Sigma$ if there exists $\alpha\in \mathcal{K}_{\infty}$ such that
	\begin{equation}\label{eq:lowerbound2}
	\alpha(\Vert h(x)-\hat h(\hat x)\Vert)\le V(x,\hat x),\quad\forall x\in X,\hat x\in\hat X,
	\end{equation}
	and for all  $x\in X,\,\hat x\in\hat X,\,\hat\nu\in\hat U$, there exists $\nu\in U$ such that
	\begin{align}\label{eq6666}
	\mathbb{E} &\Big[V(f(x,\nu,\varsigma),\hat{f}(\hat x,\hat \nu,\varsigma))\,\big|\,x,\hat{x},\nu, \hat{\nu}\Big]\leq \max\Big\{\kappa(V(x,\hat{x})),\rho_{\mathrm{ext}}(\Vert\hat\nu\Vert),\psi\Big\},\
	\end{align}
	for some $\kappa\in \mathcal{K}_\infty$ with $\kappa<\mathcal{I}_d$, $\rho_{\mathrm{ext}} \in \mathcal{K}_{\infty}\cup \{0\}$, and $\psi \in\mathbb R_{\ge 0}$.
\end{definition}
We call $\widehat\Sigma$ an abstraction of $\Sigma$, and denote by $\widehat\Sigma\preceq\Sigma$ if there exists an SSF $V$ from $\widehat\Sigma$ to $\Sigma$. 

\begin{IEEEproof}\textbf{(Proposition~\ref{proposition})}
	By defining
	\begin{align}\notag
	\mathcal{A} &= \{\Vert y_{1a_1\nu_1}(k)-y_{2a_2\nu_2}(k)\Vert<\varepsilon_1 \,|\,[a_1;a_2;a_3]\},\\\notag
	\mathcal{B} &= \{\Vert y_{2a_2\nu_2}(k)-y_{3a_3\nu_3}(k)\Vert<\varepsilon_2 \,|\,[a_1;a_2;a_3]\},\\\notag
	\mathcal{C} &= \{\Vert y_{1a_1\nu_1}(k)-y_{3a_3\nu_3}(k)\Vert<\varepsilon_1+\varepsilon_2 \,|\,[a_1;a_2;a_3]\},
	\end{align}
	we have $\mathbb{P}\{\mathcal{\bar A}\}\leq \hat\delta_1$ and  $\mathbb{P}\{\mathcal{\bar B}\}\leq \hat\delta_2,$ where $\mathcal{\bar A}$ and $\mathcal{\bar B}$ are the complement of $\mathcal{A}$ and $\mathcal{B}$, respectively. Since $\mathbb{P}\{\mathcal{A}\cap \mathcal{B}\}\leq \mathbb{P}\{\mathcal{C}\}$, we have
	\begin{align}\notag
	\mathbb{P}\{\mathcal{\bar C}\} \leq \mathbb{P}\{\mathcal{\bar A}\cup \mathcal{\bar B}\} \leq \mathbb{P}\{\mathcal{\bar A}\} + \mathbb{P}\{\mathcal{\bar B}\} \leq \hat \delta_1 + \hat \delta_2.
	\end{align}
	Then
	\begin{align}\notag
	\mathbb{P}&\left\{\sup_{0\leq k\leq T_d}\Vert y_{1a_1\nu_1}(k)-y_{3a_3\nu_3}(k)\Vert\geq\varepsilon_1+\varepsilon_2\,|\,[a_1;a_2;a_3]\right\}\leq \hat\delta_1+\hat \delta_2.
	\end{align}
\end{IEEEproof}

\begin{IEEEproof}\textbf{(Theorem~\ref{Thm: Comp})}
	We first show that for some $\mathcal{K}_\infty$ function $\alpha$, the SSF $V$ in~\eqref{Comp: Simulation Function} satisfies the inequality \eqref{eq:lowerbound2}. For any $x=[{x_1;\ldots;x_N}]\in X$ and  $\hat x=[{\hat x_1;\ldots;\hat x_N}]\in \hat X$, one gets:
	\begin{align}\notag
	\Vert h(x) - \hat h(\hat x) \Vert&= \max_i \{\Vert h_{ii}(x_i) - \hat h_{ii}(\hat x_i) \Vert\}\\\notag
	&\le\max_i \{\Vert h_{i}(x_i) - \hat h_{i}(\hat x_{i}) \Vert\}\le\max_i \{\alpha_{i}^{-1}(S_{i}( x_i, \hat x_{i}))\}\\\notag
	&\le\beta~(\max_i \{\sigma^{-1}_{i}(S_{i}(x_i, \hat x_{i}))\})=
	\beta( V(x, \hat x)),
	\end{align}
	where $\beta(s)=\max_i\Big\{\alpha^{-1}_{i}\circ \sigma_{i}(s)\Big\}$ for all $s \in \mathbb R_{\ge 0}$, which is a $\mathcal{K}_\infty$ function and~\eqref{eq:lowerbound2} holds with $\alpha=\beta^{-1}$. 
	
	We continue with showing~\eqref{eq6666}, as well. Let $\kappa(s)= \max_{i,j}\{\sigma_{i}^{-1}\circ\kappa_{ij}\circ\sigma_{j}(s)\}$. It follows from~\eqref{compositionality} that $\kappa<\mathcal{I}_d$. Since $\max_{i}\sigma_{i}^{-1}$ is concave, one can readily acquire the chain of inequalities in \eqref{Equ1b} using Jensen's inequality, and by defining $\rho_{\mathrm{ext}}$, and $\psi$ as
	\begin{IEEEeqnarray*}{rCl}
		\rho_{\mathrm{ext}}(s) &:=& \left\{\hspace{-1mm}\begin{array}{l}\max_{i}\{\sigma_{i}^{-1}\circ\rho_{\mathrm{ext}i}(s_i)\},\\
			\text{s.t.} ~~~ s_i  {\ge 0},~\|[{s_1;\ldots;s_N}]\| = s,
		\end{array}\right.\\\notag
		\psi&:=&\max_{i}\sigma_{i}^{-1}(\psi_{i}).
	\end{IEEEeqnarray*}	 
	Note that $\kappa$ and $\rho_{\mathrm{ext}}$ in \eqref{Equ1b} belong to $\mathcal{K}_\infty$ and $\mathcal{K}_\infty\cup\{0\}$, respectively, due to their definition provided above. Hence, $V$ is an SSF from $\widehat \Sigma$ to $\Sigma$ which completes the proof. 
\end{IEEEproof}

\begin{figure*}[ht!]
	\rule{\textwidth}{0.1pt}
	\begin{align}\nonumber
	\mathbb{E}&\Big[V(f(x,\nu,\varsigma),\hat{f}(\hat x,\hat \nu,\varsigma))|x,\hat x,\hat{\nu}\Big]\\\notag
	&=\mathbb{E}\Big[\max_{i}\Big\{\sigma_{i}^{-1}(S_{i}(f_i(x_i,\nu_i,w_i,\varsigma_i),\hat{f}_{i}(\hat x_{i},\hat \nu_{i},\hat w_{i},\varsigma_i)))\Big\}\,\big|\,x,\hat x,\hat{\nu}\Big]\\\notag
	&\le\max_{i}\Big\{\sigma_{i}^{-1}(\mathbb{E}\Big[S_{i}(f_i(x_i,\nu_i,w_i,\varsigma_i),\hat{f}_{i}(\hat x_{i},\hat \nu_{i},\hat w_{i},\varsigma_i))\,\big|\,x,\hat x,\hat{\nu}\Big])\Big\}\\\notag
	&=\max_{i}\Big\{\sigma_{i}^{-1}(\mathbb{E}\Big[S_{i}(f_i(x_i,\nu_i,w_i,\varsigma_i),\hat{f}_{i}(\hat x_{i},\hat \nu_{i},\hat w_{i},\varsigma_i))\,\big|\,x_i,\hat x_{i},\hat \nu_{i}\Big])\Big\}\\\notag
	&\leq\max_{i}\Big\{\sigma_{i}^{-1}(\max\{\kappa_{i}(S_{i}(x_i,\hat x_{i})),\rho_{\mathrm{int}i}(\Vert w_i-\hat w_{i}\Vert), \rho_{\mathrm{ext}i}(\Vert\hat\nu_{i}\Vert),\psi_{i}\})\Big\}\\\notag
	&=\max_{i}\Big\{\sigma_{i}^{-1}(\max\{\kappa_{i}(S_{i}(x_i,\hat x_{i})),\rho_{\mathrm{int}i}(\max_{j, j\neq i}\{\Vert w_{ij}-\hat w_{ij}\Vert\}), \rho_{\mathrm{ext}i}(\Vert\hat\nu_{i}\Vert),\psi_{i}\})\Big\}\\\notag
	&=\max_{i}\Big\{\sigma_{i}^{-1}(\max\{\kappa_{i}(S_{i}(x_i,\hat x_{i})),\rho_{\mathrm{int}i}(\max_{j, j\neq i}\{\Vert y_{ji}-\hat y_{ji}\Vert\}), \rho_{\mathrm{ext}i}(\Vert\hat\nu_{i}\Vert),\psi_{i}\})\Big\}\\\notag
	&=\max_{i}\Big\{\sigma_{i}^{-1}(\max\{\kappa_{i}(S_{i}(x_i,\hat x_{i})), \rho_{\mathrm{int}i}(\max_{j, j\neq i}\{\Vert h_j(x_j)-\hat h_{j}(
	\hat x_{j})\Vert\}), \rho_{\mathrm{ext}i}(\Vert\hat\nu_{i}\Vert),\psi_{i}\})\Big\}\\\notag
	&\leq\max_{i}\Big\{\sigma_{i}^{-1}(\max\{\kappa_{i}(S_{i}(x_i,\hat x_{i})),\rho_{\mathrm{int}i}(\max_{j , j\neq i}\{\alpha_{j}^{-1}(S_{j}(x_j, \hat x_{j}))\}),\rho_{\mathrm{ext}i}(\Vert\hat\nu_{i}\Vert),\psi_{i}\})\Big\}\\\notag
	&=\max_{i,j}\Big\{\sigma_{i}^{-1}(\max\{\kappa_{ij}(S_{j}(x_j,\hat x_{j})),\rho_{\mathrm{ext}i}(\Vert\hat\nu_{i}\Vert),\psi_{i}\})\Big\}\\\notag
	&=\max_{i,j}\Big\{\sigma_{i}^{-1}(\max\{\kappa_{ij}\circ \sigma_{j}\circ \sigma_{j}^{-1}(S_{j}(x_j,\hat x_{j})),\rho_{\mathrm{ext}i}(\Vert\hat\nu_{i}\Vert),\psi_{i}\})\Big\}\\\notag
	&\leq\max_{i,j,l}\Big\{\sigma_{i}^{-1}(\max\{\kappa_{ij}\circ \sigma_{j} \circ \sigma_{l}^{-1}(S_{l}(x_l,\hat x_{l})),\rho_{\mathrm{ext}i}(\Vert\hat\nu_{i}\Vert),\psi_{i}\})\Big\}\\\notag
	&=\max_{i,j}\Big\{\sigma_{i}^{-1}(\max\{\kappa_{ij}\circ\sigma_{j}(V(x,\hat x)),\rho_{\mathrm{ext}i}(\Vert\hat\nu_{i}\Vert),\psi_{i}\})\Big\}\\\label{Equ1b}
	&=\max\Big\{\kappa(V(x,\hat x)),\rho_{\mathrm{ext}}(\Vert\hat\nu\Vert),\psi\Big\}.
	\end{align}
	\rule{\textwidth}{0.1pt}
\end{figure*}

\begin{figure*}
	\rule{\textwidth}{0.1pt}
	\begin{align}
	&\notag\mathbb{E} \Big[S(f(x,\nu,w,\varsigma),\hat{f}(\hat x,\hat \nu,\hat w,\varsigma))\,\big|\,x,\hat{x},\hat{\nu}, w,\hat w\Big]\\\notag 
	&=(x-P\hat x)^T\Big[((A+BK)+\bar\delta(BL_1+E)F)^T M((A+BK)+\bar\delta(BL_1+E)F)\Big](x-P\hat x)\\\notag
	&~~~+2 \Big[(x-P\hat x)^T((A+BK)+\bar\delta(BL_1+E)F)^T\Big] M\Big[D(w-\hat w)\Big]+\hat \nu^T(B\tilde R-P\hat B)^T M(B\tilde R-P\hat B)\hat \nu\\\notag
	&~~~+2 \Big[(x-P\hat x)^T((A+BK)+\bar\delta(BL_1+E)F)^T\Big] M\Big[(B\tilde R-P\hat B)\hat \nu\Big]+(w-\hat w)^T D^T M D(w-\hat w)\\\notag
	&~~~+2 \Big[(w-\hat w)^T D^T\Big] M\Big[(B\tilde R-P\hat B)\hat \nu\Big]\\\notag
	&\le\begin{bmatrix}x-P\hat x\\\bar\delta F(x-P\hat x)\\\end{bmatrix}^T\!\begin{bmatrix}
	(1+2/\pi)(A\!+\!BK)^T M(A\!+\!BK) && (A+BK)^T M(BL_1+E)\\
	*&& (1+2/\pi)(B\tilde R\!-\!P\hat B)^T  M(B\tilde R\!-\!P\hat B)
	\end{bmatrix}\begin{bmatrix}x-P\hat x\\\bar\delta F(x-P\hat x)\\\end{bmatrix}\\\notag
	&~~~+p(1+2\pi+1/\pi){\Vert\sqrt{M}D\Vert_2^2}\Vert w-\hat w\Vert^2 +m(1+3\pi)\Vert\sqrt{M}(B\tilde R-P\hat B)\Vert_2^2 \Vert\hat\nu\Vert^2\\\notag
	&\le\!\begin{bmatrix}\!x\!-\!P\hat x\\\bar\delta F(x\!-\!P\hat x)\!\\\end{bmatrix}^T\!\!\!\begin{bmatrix}
	\hat\kappa M&\! \!-F^T\\
	\!-F & \!\frac{2}{b}
	\end{bmatrix}\!\!\begin{bmatrix}x\!-\!P\hat x\\\bar\delta F(x\!-\!P\hat x)\\\end{bmatrix}\!\!+\!p(1\!+\!2\pi\!+\!1/\pi){\Vert\sqrt{\!M}\!D\Vert_2^2}\Vert w\!-\!\hat w\Vert^2 \!+\!m(1\!+\!3\pi)\Vert\sqrt{\!M}(B\tilde R\!-\!P\hat B)\Vert_2^2 \Vert\hat\nu\Vert^2\\\notag
	&=\hat\kappa S(x,\hat x)\!-\!2\bar\delta(1\!\!-\!\frac{\bar\delta}{b})\!(x\!-\!P\hat x)^T\!F^T\!\!F(x\!-\!P\hat x)\!+\!p(1\!+\!2\pi\!+\!1/\pi){\Vert\!\sqrt{\!M}\!D\Vert_2^2}\Vert w\!-\!\hat w\Vert^2 \!\!+\!m(1\!+\!3\pi\!)\Vert\!\sqrt{\!M}(B\tilde R\!-\!P\hat B)\Vert_2^2 \Vert\hat\nu\Vert^2\\\notag
	&\le\hat\kappa S(x,\hat x)+p(1+2\pi+1/\pi){\Vert\sqrt{M}D\Vert_2^2}\Vert w-\hat w\Vert^2 +m(1+3\pi)\Vert\sqrt{M}(B\tilde R-P\hat B)\Vert_2^2 \Vert\hat\nu\Vert^2\\\notag
	&\leq \max\Big\{(1-(1-\tilde \pi)\tilde \kappa)(S(x,\hat{x})), (1+\tilde \delta) (\frac{1}{\tilde \kappa \tilde\pi})(p(1+2\pi+1/\pi))\Vert\sqrt{M}D\Vert_2^2\Vert w-\hat w\Vert^2,\\\label{Eq_515a}
	&~~~(1+1/\tilde \delta)(\frac{1}{\tilde \kappa \tilde\pi})(m(1+3\pi))\Vert\sqrt{M}(B\tilde R-P\hat B)\Vert_2^2 \Vert\hat\nu\Vert^2 \Big\}.
	\end{align}
	\rule{\textwidth}{0.1pt}
	\vspace{-5mm}
\end{figure*}
\begin{figure*}[ht]
	\rule{\textwidth}{0.1pt}
	\begin{align}\nonumber
	\mathbb{E}&\Big[V(f(x,\nu,\varsigma),\hat{f}(\hat x,\hat \nu,\varsigma))\,\big|\,x,\hat x,\hat{\nu}\Big]\\\notag
	&=\mathbb{E}\Big[\max_{i}\Big\{\sigma_i^{-1}(S_i(f_{i}(x_{i},\nu_{i}, w_{i},\varsigma_i),\hat{f}_i(\hat x_i,\hat \nu_i,\hat w_i,\varsigma_i)))\Big\}\,\big|\,x,\hat x,\hat{\nu}\Big]\\\notag
	&\le\max_{i}\Big\{\sigma_i^{-1}(\mathbb{E}\Big[S_i(f_{i}(x_{i},\nu_{i}, w_{i},\varsigma_i),\hat{f}_i(\hat x_i,\hat \nu_i,\hat w_i,\varsigma_i))\,\big|\, x,\hat x,\hat{\nu}\Big])\Big\}\\\notag
	&=\max_{i}\Big\{\sigma_i^{-1}(\mathbb{E}\Big[S_i(f_{i}(x_{i},\nu_{i}, w_{i},\varsigma_i),\hat{f}_i(\hat x_i,\hat \nu_i,\hat w_i,\varsigma_i))\,\big|\,x_{i},\hat x_i,\hat{\nu_i}\Big])\Big\}\\\notag
	&\leq\max_{i}\Big\{\sigma_i^{-1}(\max\{\kappa_i(S_i(x_{i},\hat x_i)),\rho_{\mathrm{int}i}(\Vert w_{i}-\hat w_i\Vert), \rho_{\mathrm{ext}i}(\Vert\hat\nu_i\Vert),\psi_i\})\Big\}\\\notag
	&=\max_{i}\Big\{\sigma_i^{-1}(\max\{\kappa_i(S_i(x_{i},\hat x_i)),\rho_{\mathrm{int}i}(\max_{j, j\neq i}\{\Vert w_{ij}-\hat w_{ij}\Vert\}), \rho_{\mathrm{ext}i}(\Vert\hat\nu_i\Vert),\psi_i\})\Big\}\\\notag
	&=\max_{i}\Big\{\sigma_i^{-1}(\max\{\kappa_i(S_i(x_{i},\hat x_i)),\rho_{\mathrm{int}i}(\max_{j, j\neq i}\{\Vert y_{ji}-\hat y_{ji}+\hat y_{ji}-\Pi_{w_{ji}}(\hat y_{ji})\Vert\}), \rho_{\mathrm{ext}i}(\Vert\hat\nu_i\Vert),\psi_i\})\Big\}\\\notag
	&\leq\max_{i}\Big\{\sigma_i^{-1}(\max\{\kappa_i(S_i(x_{i},\hat x_i)), \rho_{\mathrm{int}i}(\max_{j, j\neq i}\{\Vert h_{j}(x_{j})-\hat h_j(
	\hat x_j)\Vert+\Vert\hat y_{ji}-\Pi_{w_{ji}}(\hat y_{ji})\Vert\}), \rho_{\mathrm{ext}i}(\Vert\hat\nu_i\Vert),\psi_i\})\Big\}\\\notag
	&\leq\max_{i}\Big\{\sigma_i^{-1}(\max\{\kappa_i(S_i(x_{i},\hat x_i)),\rho_{\mathrm{int}i}(\max_{j , j\neq i}\{\alpha_j^{-1}(S_j(x_{j}, \hat x_j))+\mu_{ji}\}),\rho_{\mathrm{ext}i}(\Vert\hat\nu_i\Vert),\psi_i\})\Big\}\\\notag
	&\leq\max_{i}\Big\{\sigma_i^{-1}(\max\{\kappa_i(S_i(x_{i},\hat x_i)),\rho_{\mathrm{int}i}\circ \bar \lambda(\max_{j, j\neq i}\{\alpha_j^{-1}(S_j(x_{j}, \hat x_j))\})+\rho_{\mathrm{int}i}\circ \bar \lambda\circ(\bar \lambda-\mathcal{I}_d)^{-1}(\max_{j, j\neq i}\{\mu_{ji}\}),\\\notag
	&\quad\quad\quad\quad\rho_{\mathrm{ext}i}(\Vert\hat\nu_i\Vert),\psi_i\})\Big\}\\\notag
	&\leq\max_{i}\Big\{\sigma_i^{-1}(\max\{\kappa_i(S_i(x_{i},\hat x_i)),(\mathcal{I}_d + \tilde \delta_f)\circ\rho_{\mathrm{int}i}\circ \bar \lambda(\max_{j, j\neq i}\{\alpha_j^{-1}(S_j( x_{j}, \hat x_j))\}),\rho_{\mathrm{ext}i}(\Vert\hat\nu_i\Vert),\Lambda_i\})\Big\}\\\notag
	&=\max_{i,j}\Big\{\sigma_i^{-1}(\max\{\kappa_{ij}(S_j(x_{j},\hat x_j)),\rho_{\mathrm{ext}i}(\Vert\hat\nu_i\Vert),\Lambda_i\})\Big\}\\\notag
	&=\max_{i,j}\Big\{\sigma_i^{-1}(\max\{\kappa_{ij}\circ \sigma_j\circ \sigma_j^{-1}(S_j(x_{j},\hat x_j)),\rho_{\mathrm{ext}i}(\Vert\hat\nu_i\Vert),\Lambda_i\})\Big\}\\\notag
	&\leq\max_{i,j,l}\Big\{\sigma_i^{-1}(\max\{\kappa_{ij}\circ \sigma_j \circ \sigma_l^{-1}(S_l( x_{l},\hat x_l)),\rho_{\mathrm{ext}i}(\Vert\hat\nu_i\Vert),\Lambda_i\})\Big\}\\\notag
	&=\max_{i,j}\Big\{\sigma_i^{-1}(\max\{\kappa_{ij}\circ\sigma_j(V(x,\hat x)),\rho_{\mathrm{ext}i}(\Vert\hat\nu_i\Vert),\Lambda_i\})\Big\}\\\label{Equ11b}
	&=\max\Big\{\kappa(V(x,\hat{x})),\rho_{\mathrm{ext}}(\Vert\hat\nu\Vert),\psi\Big\}.
	\end{align}
	\rule{\textwidth}{0.1pt}
\end{figure*}

\begin{example}\label{comparable example}
	Consider the following system:
	\begin{align*}
	\Sigma:\left\{\hspace{-1.5mm}
	\begin{array}{rl}
	x_1(k+1)&\!\!\!\!=a_1x_1(k)+b_1\sqrt{\vert x_2(k)\vert} + \varsigma_1(k),\\
	x_2(k+1)&\!\!\!\!=a_2x_2(k)+b_2g(x_1(k)) ~+ \varsigma_2(k),
	\end{array}\right.
	\end{align*}
	where $0<a_1<1$, $0<a_2<1$, $b_1,b_2\in \mathbb \R$, and the function $g$ satisfies the following quadratic Lipschitz assumption: there exists an $\mathscr{L}\in\R_{>0}$ such that: $\vert g(x)-g(x')\vert\leq \mathscr{L}\vert x-x'\vert^2$ for all $x,x'\in \mathbb \R$. One can easily verify that functions $S_1(x_1,\hat x_1)=\vert x_1-\hat x_1\vert$ and $S_2(x_2,\hat x_2)=\vert x_2-\hat x_2\vert$ are stochastic pseudo-simulation functions from subsystems $x_1$ and $x_2$ to themselves, respectively. Here, one cannot come up with gain functions that globally satisfy Assumption 1 in \cite{lavaei2017compositional}. In particular, this assumption requires the existence of $\mathcal{K}_\infty$ functions being upper bounded by linear ones and lower bounded by quadratic ones which is impossible to satisfy globally. On the other hand, the proposed small-gain condition~\eqref{Assump: Kappa} is still applicable here showing that $V(x,\hat{x}):=\max\{ \sigma^{-1}_{1}\circ S_1(x_{1},\hat{x}_{1}),\sigma^{-1}_{2}\circ S_2(x_{2},\hat{x}_{2}) \}$ is a stochastic simulation function from $\Sigma$ to itself, for some appropriate $\sigma_{1},\sigma_{2} \in \mathcal{K}_{\infty}$ (with concave $\max_{1}\sigma_1^{-1}$, $\max_{2}\sigma_2^{-1}$) satisfying~\eqref{compositionality} which is guaranteed to exist if $|b_1|\sqrt{|b_2|\mathscr{L}}<1$ and $|b_2|(b_1\mathscr{L})^2<1$. Then the \emph{max} small-gain condition~\eqref{Assump: Kappa} is much more general than the \emph{classic} one proposed in~\cite{lavaei2017compositional}.
\end{example}

\begin{IEEEproof}\textbf{(Theorem~\ref{Thm_33a})}
	According to~\eqref{Eq_26a}, we have $\Vert Cx-\hat C\hat x\Vert^2\leq n\lambda_{\max}(C^TC)\Vert x- P\hat x\Vert^2$, and similarly  $\lambda_{\min}(M)\Vert x- P\hat x\Vert^2\leq(x-P\hat x)^T M(x-P\hat x)$. Then one can readily verify that  $\frac{\lambda_{\min}(M)}{n\lambda_{\max}(C^TC)}\Vert Cx-\hat C\hat x\Vert^2\le S(x,\hat x)$ holds $\forall x$, $\forall \hat x$, implying that the inequality~\eqref{Eq_2a} holds with $\alpha(s)=\frac{\lambda_{\min}(M)}{n\lambda_{\max}(C^TC)}\,s^2$ for any $s\in\mathbb R_{\geq0}$. We proceed with showing that the inequality~\eqref{Eq_3a} holds, as well. Given any $x$, $\hat x$, and $\hat \nu$, we choose $\nu$ via the following interface function:
	\begin{align}\label{Eq_40a}
	\nu=\nu_{\hat \nu}(x,\hat x,\hat \nu):=K(x-P\hat x)+Q\hat x+\tilde R\hat \nu+S\hat w+L_1\varphi(Fx)-L_2\varphi(FP \hat x),
	\end{align}
	for some matrix $\tilde R$ of an appropriate dimension. 
	By employing equations~\eqref{Eq_10a},~\eqref{Eq_15a},~\eqref{Eq_14a}, \eqref{Eq_11a}, \eqref{Eq_16a}, and also the definition of the interface function in~\eqref{Eq_40a}, we simplify
	\begin{align}\notag
	Ax+E\varphi(Fx)+B\nu(x,\hat x, \hat \nu)+Dw -P(\hat A\hat x+\hat E\varphi(\hat F\hat x)+\hat B\hat \nu+\hat D\hat w)
	+(R\varsigma-P\hat R\varsigma)
	\end{align}
	to 
	\begin{align}\label{Eq_118a}
	(A+BK)(x-P\hat x)+D(w-\hat w)+(B\tilde R-P\hat B)\hat \nu+(BL_1+E)(\varphi(Fx)-\varphi(FP\hat x)). 
	\end{align}
	From the slope restriction~\eqref{Eq_6a}, one obtains
	\begin{align}\label{Eq_119a}
	\varphi(Fx)-\varphi(FP\hat x)=\bar\delta(Fx-FP\hat x)=\bar\delta F(x-P\hat x),
	\end{align}
	where $\bar\delta$ is a function of $x$ and $\hat x$ and takes values in the interval $[0,b]$. Using~\eqref{Eq_119a}, the expression in~\eqref{Eq_118a} reduces to
	\begin{align}\notag
	((A+BK)&+\bar\delta(BL_1+E)F)(x-P\hat x)+D(w-\hat w)+(B\tilde R-P\hat B)\hat \nu. 
	\end{align}
	Using Young's inequality~\cite{young1912classes} as $cd\leq \frac{\pi}{2}c^2+\frac{1}{2\pi}d^2,$ for any $c,d\geq0$ and any $\pi>0$, and by employing Cauchy-Schwarz inequality and~\eqref{Eq_8a}, one can obtain the chain of inequalities in~\eqref{Eq_515a} in order to obtain an upper bound. Hence, the proposed $S$ in~\eqref{Eq_77a} is an SPSF from  $\widehat \Sigma$ to $\Sigma$, which completes the proof. Note that the last inequality in~\eqref{Eq_515a} is derived by applying Theorem 1 in~\cite{Abdallah2017compositional}. The functions $\alpha,\kappa\in\mathcal{K}_\infty$, and $\rho_{\mathrm{int}}$, $\rho_{\mathrm{ext}}\in\mathcal{K}_\infty\cup\{0\}$ in Definition~\ref{Def_1a} associated with $S$ in~\eqref{Eq_77a} are defined as $\alpha(s)=\frac{\lambda_{\min}(M)}{n\lambda_{\max}(C^TC)}\,s^2$, $\kappa(s):=(1-(1-\tilde \pi)\tilde \kappa)\,s$, $\rho_{\mathrm{int}}(s):=(1+\tilde \delta) (\frac{1}{\tilde \kappa \tilde\pi})(p(1+2\pi+1/\pi))\Vert\sqrt{M}D\Vert_2^2\,s^2$, $\rho_{\mathrm{ext}}(s):=(1+1/\tilde \delta)(\frac{1}{\tilde \kappa \tilde\pi})(m(1+3\pi))\Vert\sqrt{M}(B\tilde R-P\hat B)\Vert_2^2\, s^2$, $\forall s\in\mathbb R_{\ge0}$ where $\tilde \kappa = 1- \hat\kappa$, $0<\tilde \pi<1$, and $\tilde \delta> 0$. Moreover, the positive constant $\psi$ in~\eqref{Eq_3a} is equal to zero.
\end{IEEEproof}

\begin{IEEEproof}\textbf{(Theorem~\ref{Thm: Comp1})}
	We first show that the SSF $V$ in~\eqref{Comp: Simulation Function1} satisfies the inequality \eqref{eq:lowerbound2} for some $\mathcal{K}_\infty$ function $\alpha$. For any $x=[{x_1;\ldots; x_N}]\in X$ and  $\hat x=[{\hat x_1;\ldots;\hat x_N}]\in \hat X$, one gets:
	\begin{align}\notag
	\Vert h(x) - \hat h(\hat x) \Vert&= \max_i \{\Vert h_{ii}(x_{i}) - \hat h_{ii}(\hat x_i) \Vert\}\le\max_i \{\Vert h_{i}(x_{i}) - \hat h_{i}(\hat x_i) \Vert\}\\\notag
	&\le\max_i \{\alpha_i^{-1}(S_i(x_{i}, \hat x_i))\}\le\beta~(\max_i \{\sigma^{-1}_i(S_i(x_{i}, \hat x_i))\})=
	\beta( V(x,\hat x)),
	\end{align}
	where $\beta(s)=\max_i\Big\{\alpha^{-1}_i\circ \sigma_i(s)\Big\}$ for all $s \in \mathbb R_{\ge 0}$, which is a $\mathcal{K}_\infty$ function and~\eqref{eq:lowerbound2} holds with $\alpha=\beta^{-1}$. 
	
	We continue with showing~\eqref{eq6666}.
	Let $\kappa(s)= \max_{i,j}\{\sigma_i^{-1}\circ\kappa_{ij}\circ\sigma_j(s)\}$. It follows from~\eqref{compositionality1} that $\kappa<\mathcal{I}_d$. Since $\max_{i}\sigma_i^{-1}$ is concave, one can readily get the chain of inequalities in \eqref{Equ11b} using Jensen's inequality, the inequality~\eqref{eq:Pi_mu}, and by defining $\rho_{\mathrm{ext}}(\cdot)$, and $\psi$ as
	\begin{IEEEeqnarray*}{rCl}
		\rho_{\mathrm{ext}}(s) &:=& \left\{\hspace{-1.5mm}\begin{array}{l}\max_{i}\{\sigma_i^{-1}\circ\rho_{\mathrm{ext}i}(s_i)\},\\
			\text{s.t.} ~~~ s_i  {\ge 0},~\|[{s_1;\ldots;s_N}]\| = s,
		\end{array}\right.\\\notag
		\psi&:=&\max_{i}\sigma_i^{-1}(\Lambda_i),
	\end{IEEEeqnarray*} 
	where $\Lambda_i: =(\mathcal{I}_d + \tilde \delta_f^{-1})\circ (\rho_{\mathrm{int}i}\circ \bar \lambda\circ(\bar \lambda-\mathcal{I}_d)^{-1}(\max_{j, j\neq i}\{\mu_{ji}\})+\psi_i)$.
	Hence, $V$ is an SSF from $\widehat \Sigma$ to $\Sigma$ which completes the proof. 
\end{IEEEproof}
\begin{remark}
Note that to show Theorem~\ref{Thm: Comp1}, we have employed the following inequalities:
\begin{equation}\notag
\left\{\hspace{-1.5mm}\begin{array}{l}\rho_{\mathrm{int}}(a+b)\leq\rho_{\mathrm{int}}\circ\bar \lambda(a)+\rho_{\mathrm{int}}\circ\bar \lambda\circ(\bar \lambda- \mathcal{I}_d)^{-1}(b), \\
a+b\leq\max\{(\mathcal{I}_d+\tilde \delta_f)(a),(\mathcal{I}_d+\tilde \delta_f^{-1})(b)\},\\
\end{array}\right.
\end{equation}
for any  $a,b\in\mathbb R_{\ge 0}$, where $\rho_{\mathrm{int}}, \tilde \delta_f, \bar \lambda, (\bar \lambda- \mathcal{I}_d)\in\mathcal{K}_\infty$. 
\end{remark}

\begin{IEEEproof}\textbf{(Theorem~\ref{Thm_5a})}
		Given the Lipschitz assumption on $h$, since $\Sigma$ is incrementally input-to-state stable, and from \eqref{Con555}, $\forall x\in X$ and $ \forall \hat x \in \hat X
	$, we get 
	\begin{align}\notag
	\Vert h(x)-\hat h(\hat x ) \Vert \leq \tilde{\alpha}(\Vert x-\hat x\Vert)\leq \hat{\alpha}(S(x,\hat x)),
	\end{align}
	where $\hat \alpha = \tilde \alpha\circ \underline \alpha^{-1}$, which satisfies \eqref{Eq_2a} with  $\alpha(s) \Let \hat{\alpha}^{-1}(s) $ $\forall s\in \R_{\geq0}$.
	Now by taking the conditional expectation from \eqref{Eq65}, $\forall x  \in X, \forall \hat x  \in \hat X, \forall \hat \nu\in \hat U,\forall w\in W,\forall \hat w \in \hat W$, we have 
	\begin{align}\notag
	\mathbb{E}&\Big[S(f(x,\hat{\nu},w,\varsigma),\hat f(\hat{x},\hat{\nu},\hat{w},\varsigma))\,\big|\,x,\hat x,\hat \nu, w, \hat w\Big]\\\notag
	&-\mathbb{E}\Big[S(f(x,\hat{\nu},w,\varsigma),f(\hat{x},\hat{\nu},\hat{w},\varsigma))\,\big|\,x,\hat x,\hat \nu,w,\hat w\Big]\\\notag
	&\leq\mathbb{E}\Big[\gamma(\Vert\hat f(\hat{x},\hat{\nu},\hat{w},\varsigma)-f(\hat{x},\hat{\nu},\hat{w},\varsigma)\Vert)\,\big|\,x,\hat x,\hat \nu, w, \hat w\Big],
	\end{align}
	where $\hat f(\hat{x},\hat{\nu},\hat{w},\varsigma) = \Pi_x(f(\hat{x},\hat{\nu},\hat{w},\varsigma))$. Using~\eqref{eq:Pi_delta}, the above inequality reduces to
	\begin{align}\notag
	\mathbb{E}&\Big[S(f(x,\hat{\nu}, w,\varsigma),\hat f(\hat{x},\hat{\nu},\hat{w},\varsigma))\,\big|\,x,\hat x,\hat \nu, w, \hat w\Big]\\\notag
	&-\mathbb{E}\Big[S(f(x,\hat{\nu},w,\varsigma),f(\hat{x},\hat{\nu},\hat{w},\varsigma))\,\big|\,x,\hat x,\hat \nu,w,\hat w\Big]\leq\gamma(\delta).
	\end{align}
	
	Employing \eqref{Con854}, we get 
	\begin{align}\notag
	\mathbb{E}&\Big[S(f(x,\hat{\nu},w,\varsigma),f(\hat{x},\hat{\nu},\hat{w},\varsigma))\,\big|\,x,\hat x,\hat \nu,w,\hat w\Big]\leq S(x,\hat x)-\bar{\kappa}(S(x,\hat x))+\bar \rho_{\mathrm{int}}(\Vert w-\hat w\Vert).
	\end{align}
	
	It follows that $\forall x \in X, \forall \hat x \in \hat X, \forall \hat u \in \hat U,$ and $\forall w \in W,\forall \hat w \in \hat W $,
	\begin{align}\notag
	\mathbb{E}&\Big[S(f(x,\hat{\nu}, w,\varsigma),\hat f(\hat{x},\hat{\nu},\hat{w},\varsigma))\,\big|\,x,\hat x,\hat \nu, w, \hat w\Big]-S(x,\hat{x})\\\notag
	&\leq-\bar{\kappa}(S(x,\hat{x}))+\bar \rho_{\mathrm{int}}(\Vert w-\hat w\Vert)+\gamma(\delta).
	\end{align}
	Using the previous inequality and by employing the similar argument as the one in~\cite[Theorem 1]{Abdallah2017compositional}, one obtains
	\begin{align}\notag
	\mathbb{E}&\Big[S(f(x,\hat{\nu},w,\varsigma),\hat f(\hat{x},\hat{\nu},\hat{w},\varsigma))\,\big|\,x,\hat x,\hat \nu, w, \hat w\Big]\\\label{general setting}
	&\leq\max\Big\{\tilde{\kappa}_f(S(x,\hat{x})),\tilde \rho_{\mathrm{int}}(\Vert w-\hat w\Vert),\tilde \gamma(\delta)\Big\},
	\end{align}
	where $\tilde {\kappa}_f =\mathcal{I}_d -(\mathcal{I}_d - \tilde {\pi}_f )\circ \underline \kappa$, $\tilde \rho_{\mathrm{int}}=(\mathcal{I}_d + \tilde {\delta}_f)\circ\underline \kappa^{-1}\circ\tilde {\pi}_f  ^{-1}\circ\bar\lambda\circ\bar \rho_{\mathrm{int}}$, $\tilde \gamma=(\mathcal{I}_d + \tilde {\delta}_f^{-1})\circ\underline \kappa^{-1}\circ\tilde {\pi}_f  ^{-1}\circ\bar\lambda\circ(\bar\lambda - \mathcal{I}_d)^{-1}\circ\gamma$ where $\tilde {\delta}_f, \tilde {\pi}_f ,\bar\lambda, \underline \kappa $ are some arbitrarily chosen $\mathcal{K}_\infty$ functions with $\mathcal{I}_d - \tilde {\pi}_f  \in \mathcal{K}_\infty$, $\bar\lambda - \mathcal{I}_d \in \mathcal{K}_\infty$, $\mathcal{I}_d - \underline \kappa \in \mathcal{K}_\infty$, and $\underline \kappa \leq \bar \kappa$. Then the inequality \eqref{Eq_3a} is satisfied
	with $\nu=\hat{\nu}$, $\kappa=\tilde{\kappa}_f$, $ \rho_{int} = \tilde \rho_{int}$, and $\rho_{ext}\equiv 0$, and $\psi=\tilde \gamma(\delta)$. Hence $S$ is an SPSF from $\widehat \Sigma$ to $\Sigma$. 
\end{IEEEproof}

\begin{IEEEproof}\textbf{(Theorem~\ref{Thm_3a})}
	Since $\hat C =C$, we have $\Vert C x-\hat C\hat x\Vert^2\leq n\lambda_{\max}(C^TC)\Vert x- \hat x\Vert^2$, and similarly  $\lambda_{\min}(M)\Vert x- \hat x\Vert^2\leq(x-\hat x)^T M(x-\hat x)$. One can readily verify that  $\frac{\lambda_{\min}(M)}{n\lambda_{\max}(C^TC)}\Vert C x-\hat C\hat x\Vert^2\le S(x,\hat x)$ holds $\forall x$, $\forall \hat x$, implying that the inequality~\eqref{Eq_2a} holds with $\alpha(s)=\frac{\lambda_{\min}(M)}{n\lambda_{\max}(C^TC)}\,s^2$ for any $s\in\mathbb R_{\geq0}$. We proceed with showing that the inequality~\eqref{Eq_3a} holds, as well. Given any $ x$, $\hat x$, and $\hat \nu$, we choose $\nu$ via the following interface function:
	\begin{align}\label{Eq_255}
	\nu=\nu_{\hat \nu}(x,\hat x,\hat \nu):=K(x-\hat x)+\hat \nu.
	\end{align}
	By employing the definition of the interface function, we simplify
	\begin{align}\notag
	Ax +E\varphi(Fx)+ B\nu(x,\hat x, \hat \nu) + Dw +R\varsigma-\Pi_x(A\hat x + E\varphi(F\hat x)+B\hat \nu +D\hat w+ R\varsigma)
	\end{align}
	to 
	\begin{align}\label{Eq_552}
	(A+BK)(x-\hat x)+ D(w-\hat w) +E(\varphi(Fx)-\varphi(F\hat x)) +\bar N,
	\end{align}
	where $\bar N = A\hat x +E\varphi(F\hat x)+B\hat \nu+ D\hat w +  R\varsigma-\Pi_x(A\hat x+E\varphi(F\hat x)+ B\hat \nu + D\hat w + R\varsigma)$.
	From the slope restriction~\eqref{Eq_6a}, one obtains
	\begin{align}\label{Eq_129a}
	\varphi(Fx)-\varphi(F\hat x)=\bar\delta(Fx-F\hat x)=\bar\delta F(x-\hat x),
	\end{align}
	where $\bar\delta$ is a function of $x$ and $\hat x$ and takes values in the interval $[0,b]$. Using~\eqref{Eq_129a}, the expression in~\eqref{Eq_552} reduces to
	\begin{align}\notag
	((A+BK)+\bar\delta EF)(x-\hat x)+D(w-\hat w)+\bar N.
	\end{align}
	Using Young's inequality~\cite{young1912classes}, Cauchy-Schwarz inequality and~\eqref{Eq_88a}, and since 
	\begin{align}\notag
	\left\{\hspace{-1.5mm}\begin{array}{l}\Vert \bar N\Vert~\leq~ \delta,\\
	\bar N^T M \bar N \leq n\lambda_{\max}(M)\delta^2,\end{array}\right.
	\end{align}
	one can obtain the chain of inequalities in~\eqref{Eq_305a}. Hence, the proposed $S$ in~\eqref{Eq_7a} is an SPSF from  $\widehat \Sigma$ to $\Sigma$, which completes the proof. Note that the last inequality in~\eqref{Eq_305a} is derived by applying Theorem 1 in~\cite{Abdallah2017compositional}. The functions $\alpha,\kappa\in\mathcal{K}_\infty$, and $\rho_{\mathrm{int}}$, $\rho_{\mathrm{ext}}\in\mathcal{K}_\infty\cup\{0\}$ in Definition~\ref{Def_1a} associated with $S$ in~\eqref{Eq_7a} are defined as $\alpha(s)=\frac{\lambda_{\min}(M)}{n\lambda_{\max}(C^TC)}\,s^2$, $\kappa(s):=(1-(1-\tilde \pi)\tilde \kappa)\,s$, $\rho_{\mathrm{int}}(s):=(1+\tilde \delta) (\frac{1}{\tilde \kappa \tilde\pi})(p(1+2\pi+1/\pi))\Vert\sqrt{M}D\Vert_2^2\,s^2$, $\rho_{\mathrm{ext}}(s):=0$, $\forall s\in\mathbb R_{\ge0}$ where $\tilde \kappa = 1- \hat\kappa$, $0<\tilde \pi <1$, and $\tilde \delta > 0$. Moreover, the positive constant $\psi$ in~\eqref{Eq_3a} is $\psi=(1+1/\tilde \delta)(\frac{1}{\tilde \kappa \tilde\pi})(n(1+3\pi)\lambda_{\max}{(M))}\,\delta^2$.	
\end{IEEEproof}

\begin{figure*}[ht!]
	\rule{\textwidth}{0.1pt}
	\begin{align}
	\notag\mathbb{E}& \Big[S(f(x,\nu, w,\varsigma),\hat{f}(\hat x,\hat \nu,\hat w,\varsigma))\,\big|\,x,\hat{x},\hat{\nu}, w,\hat w\Big]\\\notag 
	&=(x\!-\!\hat x)^T\Big[((A\!+\!BK)\!+\!\bar\delta EF)^T M((A\!+\!BK)\!+\!\bar\delta EF)\Big](x\!-\!\hat x)+2 \Big[(x\!-\!\hat x)^T((A\!+\!BK)\!+\!\bar\delta EF)^T\Big] M\Big[D(w\!-\!\hat w)\Big]\\\notag
	&~~~+2 \Big[(x-\hat x)^T((A+BK)+\bar\delta EF)^T\Big] M\mathbb{E}\Big[\bar N\,\big|\,x,\hat x, \hat \nu,w, \hat w\Big]+2 \Big[(w-\hat w)^T D^T\Big] M \mathbb{E}\Big[\bar N\,\big|\,x,\hat x, \hat \nu,w, \hat w\Big]\\\notag
	&~~~+(w-\hat w)^T D^T MD(w-\hat w)+\mathbb{E}\Big[\bar N^T M \bar N\,\big|\,x,\hat x, \hat \nu,w, \hat w\Big]\\\notag
	&\le\begin{bmatrix}x\!-\!\hat x\\\bar\delta F(x\!-\!\hat x)\\\end{bmatrix}^T\begin{bmatrix}
	(1+2/\pi)(A+BK)^T M(A+BK) && (A+BK)^T M E\\
	*&& (1+2/\pi)E^T M E
	\end{bmatrix}\begin{bmatrix}x\!-\!\hat x\\\bar\delta F(x\!-\!\hat x)\\\end{bmatrix}\\\notag
	&~~~+p(1\!+\!2\pi\!+\!1/\pi){\Vert\sqrt{M}D\Vert_2^2}\Vert w-\hat w\Vert^2 +n(1+3\pi)\lambda_{\max}{(M)}\,\delta^2\\\notag
	&\le\begin{bmatrix}x\!-\!\hat x\\\bar\delta F(x\!-\!\hat x)\\\end{bmatrix}^T\begin{bmatrix}
	\hat\kappa M&\!\! -F^T\\
	-F & \!\!\frac{2}{b}
	\end{bmatrix}\begin{bmatrix}x\!-\!\hat x\\\bar\delta F(x\!-\!\hat x)\\\end{bmatrix}\!+\!p(1\!+\!2\pi\!+\!1/\pi){\Vert\sqrt{M}D\Vert_2^2}\Vert w\!-\!\hat w\Vert^2 \!+\!n(1\!+\!3\pi)\lambda_{\max}{(M)}\,\delta^2\\\notag
	&=\hat\kappa S(x,\hat x)\!-\!2\bar\delta(1\!\!-\!\frac{\bar\delta}{b})(x\!-\!\hat x)^TF^T F( x\!-\!\hat x)\!+\!p(1\!+\!2\pi\!+\!1/\pi){\Vert\sqrt{M}D\Vert_2^2}\Vert w\!-\!\hat w\Vert^2 \!+\! n(1\!+\!3\pi)\lambda_{\max}{(M)}\,\delta^2\\\notag
	&\le\hat\kappa S(x,\hat x)+(p(1\!+\!2\pi\!+\!1/\pi))\Vert\sqrt{M}D\Vert_2^2\Vert w-\hat w\Vert^2 +n(1+3\pi)\lambda_{\max}{(M)}\,\delta^2\\\notag
	& \leq \!\max\!\Big\{(1\!-\!(1\!-\!\tilde \pi)\tilde \kappa)(S(x,\hat x)), (1\!+\!\tilde \delta) (\frac{1}{\tilde \kappa \tilde\pi})(p(1\!+\!2\pi\!+\!1/\pi))\Vert\sqrt{M} D\Vert_2^2\Vert w\!-\!\hat w\Vert^2\!,\\\label{Eq_305a}
	&\quad\quad\quad\quad\!\!(1\!+\!1/\tilde \delta)(\frac{1}{\tilde \kappa \tilde\pi})(n(1\!+\!3\pi)\lambda_{\max}{(M))}\,\delta^2 \Big\}.
	\end{align}
	\rule{\textwidth}{0.1pt}
\end{figure*}

\end{document}